\documentclass[10pt,aps,prd,twocolumn,superscriptaddress,amsmath, nofootinbib]{revtex4-1}

\usepackage{aas_macros}
\usepackage{tikz}
\usepackage{layouts}
\usepackage{colortbl}
\usepackage{lipsum}
\usepackage[hidelinks]{hyperref}
\usepackage[capitalize]{cleveref}
\Crefname{equation}{Equation}{Equations}
\crefname{equation}{Eq.}{Eqs.}
\Crefname{figure}{Figure}{Figures}
\crefname{figure}{Fig.}{Figs.}
\Crefname{table}{Table}{Tables}
\crefname{table}{Tab.}{Tabs.}
\Crefname{section}{Section}{Sections}
\crefname{section}{Sec.}{Secs.}

\newcommand{\LCDM}{$\Lambda$CDM }
\newcommand{\Planck}{\textit{Planck}}
\newcommand{\DES}{DES}
\newcommand{\SDSS}{BOSS}
\newcommand{\SHOES}{S$H_0$ES}
\newcommand{\CosmoChord}{\texttt{CosmoChord}}

\newcommand{\CosmoMC}{\texttt{CosmoMC}}
\renewcommand{\d}[2][]{\operatorname{d}^{#1}\!{#2}}

\begin{document}
\title{Quantifying dimensionality: Bayesian cosmological model complexities}
\author{Will Handley}
\email[]{wh260@mrao.cam.ac.uk}
\affiliation{Astrophysics Group, Cavendish Laboratory, J.J.Thomson Avenue, Cambridge, CB3 0HE, UK}
\affiliation{Kavli Institute for Cosmology, Madingley Road, Cambridge, CB3 0HA, UK}
\affiliation{Gonville \& Caius College, Trinity Street, Cambridge, CB2 1TA, UK}

\author{Pablo Lemos}
\email[]{pablo.lemos.18@ucl.ac.uk}
\affiliation{Department of Physics and Astronomy, University College London, Gower Street, London, WC1E 6BT, UK}

\begin{abstract}
    We demonstrate a measure for the effective number of parameters constrained by a posterior distribution in the context of cosmology. In the same way that the mean of the Shannon information (i.e.\ the Kullback-Leibler divergence) provides a measure of the strength of constraint between prior and posterior, we show that the variance of the Shannon information gives a measure of dimensionality of constraint. We examine this quantity in a cosmological context, applying it to likelihoods derived from the cosmic microwave background, large scale structure and supernovae data. We show that this measure of Bayesian model dimensionality compares favourably both analytically and numerically in a cosmological context with the existing measure of model complexity used in the literature.
\end{abstract}

\maketitle

\section{Introduction}\label{sec:introduction}

With the development of increasingly complex cosmological experiments, there has been a pressing need to understand model complexity in cosmology over the last few decades. The \LCDM{} model of cosmology is surprisingly efficient in its parameterisation of the background Universe and its fluctuations, needing only six parameters to successfully describe individual observations from all cosmological datasets~\cite{lcdm}. However, different observational techniques constrain distinct combinations of these parameters. In addition, the systematic effects that affect various observations introduce a large number of additional nuisance parameters; around twenty in both the analyses of the Dark Energy Survey~\cite{DESParameters2017} and \Planck{} collaborations~\cite{PlanckParameters2018}. 

These nuisance parameters are not always chosen in an optimal way from the point of view of sampling, with known degeneracies between each other and with the cosmological parameters. This complicates quantifying the effective number of parameters constrained by the data. Examples of these parameter degeneracies are the degeneracy between the amplitude of the primordial power spectrum $A_s$ and the optical depth to reionisation $\tau$ in the combination $A_s e^{- 2 \tau}$ in temperature anisotropies of the CMB, or the degeneracy between the intrinsic alignment amplitude and the parameter combination $S_8 \equiv \sigma_8 (\Omega_m / 0.3)^{0.5}$ in cosmic shear measurements, where $\Omega_m$ is the present-day matter density, and $\sigma_8$ is the present-day linear root-mean-square amplitude of the matter power spectrum~\cite{Troxel2015,Joachimi2015,Efstathiou2018}.

Quantifying model complexity is important beyond increasing our understanding of the data. It is necessary to measure the effective number of constrained parameters to quantify tension between datasets. The authors found this in \citet{Tension}. The pre-print version of~\cite{Tension} used the Bayesian Model Complexity (BMC) introduced in \citet{Spiegelhalter}, which the authors found unsatisfactory. Motivated by this, in this work we examine an improved {\em Bayesian model dimensionality\/} (BMD) to quantify the effective number of dimensions constrained by the data.   
Whilst the BMD measure has been introduced in the past by numerous authors~\cite{gelman,Spiegelhalter2,Skilling2006, Raveri2016, Raveri2019,Kunz,information_criteria}, in this work we provide novel interpretations in terms of information theory, and compare its performance with the BMC in a modern numerical cosmological context.

In \cref{sec:background} we introduce the notation and mathematical formalism, and some of the relevant quantities such as the Bayesian Evidence, the Shannon information and the Kullback-Leibler divergence. We also discuss some of the problems associated with Principal Component Analyses (PCA), that have been used to quantify model complexity in cosmology in the past. 

In \cref{sec:dimensionality} we discuss dimensionality in a Bayesian framework, describing the Bayesian model complexity of \citet{Spiegelhalter}, and introducing the Bayesian model dimensionality. We explain the usage of model dimensionality in the context of some analytical examples. 
Finally, in \cref{sec:numerics}, we apply Bayesian model dimensionality to real data, using four different cosmological datasets. 
We summarise our conclusions in \cref{sec:conclusion}.

\section{Background}\label{sec:background}
\begin{figure*}
	\includegraphics[]{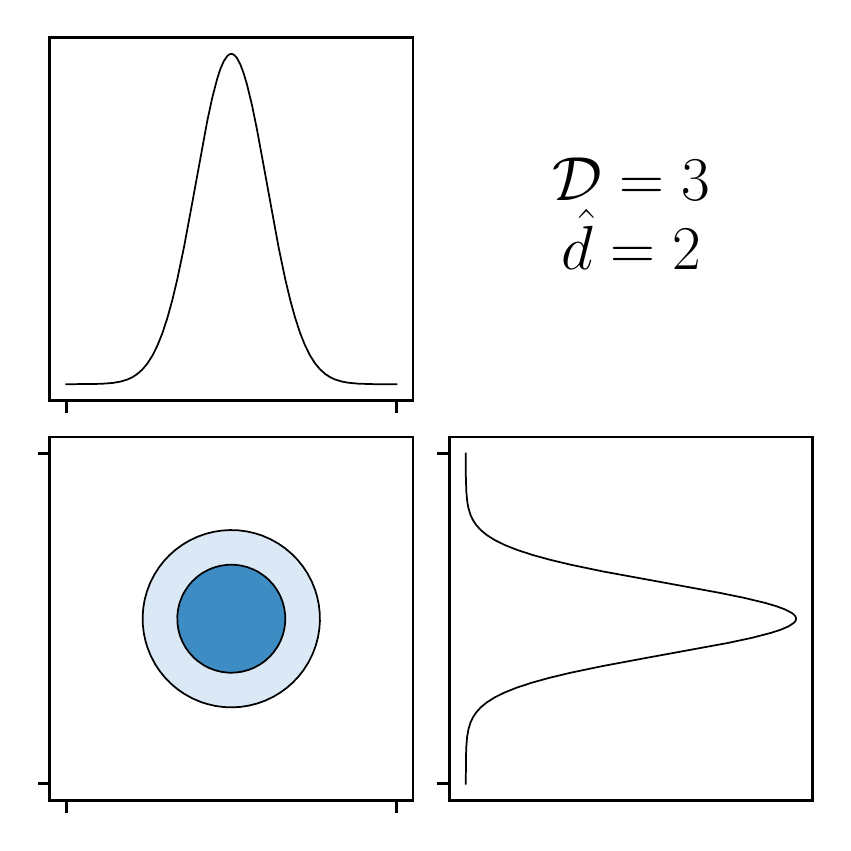}
    \hfill
	\includegraphics[]{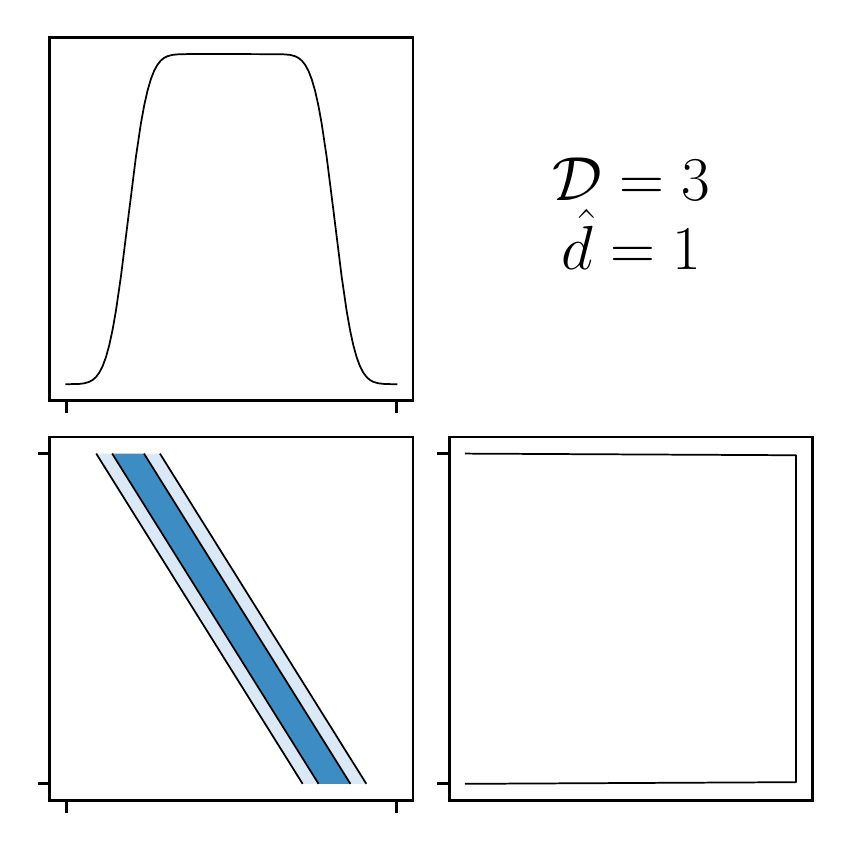}
	\caption{%
        Distributions with the same Kullback-Leibler divergence, but differing dimensionalities. Both the right and left-hand plots indicate two-dimensional probability distributions. In each plot, the lower left panel is a two-dimensional contour plot indicating the iso-probability contours enclosing $66\%$ and $95\%$ of the probability mass. The upper and lower right panels indicate the one-dimensional marginal probability distributions. There is an implicit uniform prior over the ranges indicated by the axis ticks.\label{fig:dimensions}
	}
\end{figure*}

In this section we establish notation and introduce the key inference quantities used throughout this paper. For a more detailed account of Bayesian statistics, the reader is recommended the paper by~\citet{Trotta2008}, or the text books by \citet{MacKay2002} and \citet{Sivia}.

\subsection{Bayes theorem}

In the context of Bayesian inference, a predictive model $\mathcal{M}$ with free parameters $\theta$ can use data $D$ to both provide constraints on the model parameters and infer the relative probability of the model via Bayes theorem
\begin{align}
    P(D|\theta)\times P(\theta) &= P(\theta|D)\times P(D), \\
    \mathcal{L} \times \pi &= \mathcal{P} \times \mathcal{Z},\label{eqn:bayes}
\end{align}
which should be read as ``likelihood times prior is posterior times evidence''. Whilst traditionally Bayes' theorem is rearranged to in terms of the posterior $\mathcal{P}=\mathcal{L}\pi/\mathcal{Z}$, \cref{eqn:bayes} is the form preferred by Skilling~\cite{Skilling2006}, and has since been used by other cosmologists~\cite{Hannestad}. In Skilling's form it emphasises that the inputs to inference are the model, defined by the likelihood and the prior, whilst the outputs are the posterior and evidence, used for parameter estimation and model comparison respectively.

\subsection{Shannon information}

The Shannon information~\cite{Shannon:1949} is defined as
\begin{equation}
    \mathcal{I}(\theta) = \log \frac{\mathcal{P}(\theta)}{\pi(\theta)},
    \label{eqn:shannon}
\end{equation}
and is also known as the information content, self-information or surprisal of $\theta$. The Shannon information represents the amount of information gained in nats (natural bits) about $\theta$ when moving from the prior to the posterior.

The Shannon information has the fundamental property that for independent parameters the information is additive
\begin{align}
    \mathcal{P}(\theta_1,\theta_2)&=\mathcal{P}_1(\theta_1)\mathcal{P}_2(\theta_2), \nonumber\\
    \pi(\theta_1,\theta_2)&=\pi_1(\theta_1)\pi_2(\theta_2), \nonumber\\
    \Rightarrow \mathcal{I}(\theta_1,\theta_2)&=\mathcal{I}_1(\theta_1)+\mathcal{I}_2(\theta_2)
\end{align}
Indeed it can be easily shown that the property of additivity defines \cref{eqn:shannon} up to the base of the logarithm: i.e.\ if one wishes to define a measure of information provided by a posterior that is additive for independent parameters, then one is forced to use \cref{eqn:shannon}. Additivity is an important concept used throughout this paper, as it forms the underpinning of a measurable quantity. For more detail, see Skilling's chapter in \cite{bayesian_methods}.

\subsection{Kullback-Leibler divergence}

The Kullback-Leibler divergence~\cite{Kullback:1951} is defined as the average Shannon information over the posterior
\begin{equation}
    \mathcal{D} = \int \mathcal{P}(\theta)\log\frac{\mathcal{P}(\theta)}{\pi(\theta)}\d{\theta} = \left\langle\log \frac{\mathcal{P}}{\pi}\right\rangle_{\mathcal{P}} = \left\langle\mathcal{I}\right\rangle_{\mathcal{P}}
    \label{eqn:kullback}
\end{equation}
and therefore quantifies in a Bayesian sense how much information is provided by the data $D$. Since the Shannon information is defined relative to the prior, the Kullback-Leibler divergence naturally has a strong prior dependency~\cite{Tension}. It has been widely utilised in cosmology~\cite{Hoyosa:2004,Verde:2013,Seehars2014,Seehars2016,Grandis2016,Raveri2016,HthreeL,Grandis2016b,Zhao2017,Nicola2017,Nicola2019} for a variety of analyses.

Since the Kullback-Leibler divergence is a linear function of the Shannon information, $\mathcal{D}$ is also measured in nats and is an additive quantity for independent parameters. 

Posterior averages such as \cref{eqn:kullback} in some cases can be numerically computed using samples generated by techniques such as Metropolis-Hastings~\cite{MH}, Gibbs Sampling~\cite{Gibbs} or Hamiltonian Monte Carlo~\cite{HMC}. However, computation of the Kullback-Leibler divergence is numerically more challenging, since it requires knowledge of normalised posterior densities $\mathcal{P}$, or equivalently a computation of the evidence $\mathcal{Z}$, which requires more intensive techniques such as nested sampling~\cite{Skilling2006}.

\subsection{Bayesian model complexity}

{%
\begin{table}
\begin{tabular}{|c|c|c||c|c|c|c|}
\hline
Likelihood                  & $d_{\rm Cosmo}$ & $d_{\rm Nuis}$ & $d_{\rm Total}$ \\
\hline
\SHOES{}                 & $6$ & $0$ & $6$ \\
\SDSS{}                  & $6$ & $0$ & $6$ \\ 
\DES{}                   & $6$ & $20$ & $26$ \\ 
\Planck{}                & $6$ & $15$ & $21$ \\
\hline
\end{tabular}
\caption{Number of parameters sampled over in cosmological likelihoods. $d_{\rm Cosmo}$ is the number of cosmological parameters, $d_{\rm Nuis}$ is the number of nuisance parameters, and $d_{\rm Total} = d_{\rm Cosmo} + d_{\rm Nuis}$ is the total number. Note that we sample over the same six cosmological parameters for all likelihoods, even though we know that some likelihoods cannot constrain certain parameters. For the combinations of two likelihoods, the total number is $d_{\rm Total}^{A,B} = d_{\rm Cosmo} + d_{\rm Nuis}^A + d_{\rm Nuis}^B$. \label{tab:numparams}}
\end{table}
}

Whilst the Kullback-Leibler divergence provides a well-defined measure of the overall compression from posterior to prior, it marginalises out any individual parameter information. As such, $\mathcal{D}$ tells us nothing of which parameters are providing us with information, or equally how many parameters are being constrained by the data. 

As a concrete example, consider the two posteriors illustrated in \cref{fig:dimensions}. In this case, both distributions have the same Kullback-Leibler divergence, but give very different parameter constraints. For the first distribution, both parameters are well constrained. In the second distribution, the one-dimensional marginal distributions show that the first parameter is slightly constrained, whilst the second parameter is completely unconstrained and identical to the prior. The full two-dimensional distribution tells a different story, showing that both parameters are heavily correlated, and that there is a strong constraint on a specific combination of parameters. In reality this is therefore a one-dimensional constraint that has been garbled across two parameters.

For the two-dimensional case in \cref{fig:dimensions} we can by eye determine the number of constrained parameters, but in practical cosmological situations this is not possible. The cosmological parameter space of $\Lambda$CDM is six- (arguably seven-)dimensional~\cite{lcdm}, and modern likelihoods introduce a host of nuisance parameters to combat the influence of foregrounds and systematics. For example the Planck likelihood~\cite{PlanckLikelihoods2015} is in total 21-dimensional, the \DES{} likelihood~\cite{DESParameters2017} is 26-dimensional, and their combination 41-dimensional (\cref{tab:numparams}). Whilst samples from the posterior distribution represent a near lossless compression of the information present in this distribution, it goes without saying that visualising a 40-dimensional object is challenging. Triangle/corner plots~\cite{corner} represent marginalised views of this information and can hide hidden correlations and constraints between three or more parameters. The fear is that one could misdiagnose a dataset that has powerful constraints if \cref{fig:dimensions} occurred in higher dimensions. It would be helpful if there were a number $d$ similar to the Kullback-Leibler divergence $\mathcal{D}$ which quantifies the effective number of constrained parameters.

To this end, \citet{Spiegelhalter,Spiegelhalter2} introduced the Bayesian model complexity, defined as
\begin{align}
    \frac{\hat{d}}{2} &= \log \frac{P(\hat\theta)}{\pi(\hat\theta)} - \left\langle \log \frac{\mathcal{P}}{\pi}\right\rangle_\mathcal{P}\nonumber\\
    &= \mathcal{I}(\hat{\theta}) - \left\langle \mathcal{I}\right\rangle_\mathcal{P}
    \label{eqn:bmc}
\end{align}
In this case, the model complexity measures the difference between the information at some point $\hat{\theta}$ and the average amount of information. It thus quantifies how much overconstraint there is at $\hat{\theta}$, or equivalently the degree of model complexity. This quantity been historically used in several cosmological analyses~\cite{Kunz,Bridges,Tension,Raveri2019}.

There is a degree of arbitrariness in \cref{eqn:bmc} via the choice of point estimator $\hat\theta$. Typical recommended choices include the posterior mean
\begin{equation}
    \hat{\theta}_\mathrm{m} = 
    \int \theta \mathcal{P}(\theta)\d{\theta} = \left\langle\theta\right\rangle_{\mathcal{P},}
    \label{eqn:theta_mean}
\end{equation}
the posterior mode
\begin{equation}
    \hat{\theta}_\mathrm{mp} = \max_{\theta} \mathcal{P}(\theta) 
    \label{eqn:theta_mp}
\end{equation}
or the maximum likelihood point
\begin{equation}
    \hat{\theta}_\mathrm{ml} = \max_{\theta} \mathcal{L}(\theta) = \max_{\theta}  \mathcal{I}(\theta). 
    \label{eqn:theta_ml}
\end{equation}

For the multivariate Gaussian case, $\hat{d}$ coincides with the actual dimensionality $d$ for all three of these estimators. 

Unlike the Kullback-Leibler divergence, the BMC is only weakly prior dependent, since the evidence contributions in \cref{eqn:bmc} cancel
\begin{equation}
    \hat{d} = 2\log \mathcal{L}(\hat{\theta}) - \left\langle2\log\mathcal{L}\right\rangle_\mathcal{P}.
\end{equation}
The model dimensionality thus only changes with prior $\pi$ if the posterior bulk is significantly altered by changing the prior. For example $\hat{d}$ does not change if one merely expands the widths of a uniform prior that encompasses the posterior (in contrast to the evidence and Kullback-Leibler divergence).

Finally, the model complexity in \cref{eqn:bmc} has the advantage of an information-theoretic backing and, like the Shannon information and Kullback-Leibler divergence, is additive for independent parameters.

\subsection{The problem with principle component analysis}

Intuitively from \cref{fig:dimensions} one might describe the distribution as having one ``component'' that is well constrained, and another component for which the posterior provides no information.

The approach that is then followed by many researchers is to perform a principle component analysis (PCA), which proceeds thus
\begin{enumerate}
    \item Compute the posterior covariance matrix
        \begin{equation}
            \Sigma = \left\langle (\theta-\bar\theta)(\theta-\bar\theta)^T \right\rangle_{\mathcal{P}}, \qquad \bar{\theta} = \left\langle \theta \right\rangle_\mathcal{P}
            \label{eqn:posterior_covariance}
        \end{equation}
    \item Compute the real eigenvalues $\lambda^{(i)}$ and eigenvectors $\Theta^{(i)}$ of $\Sigma$, defined via the equation
        \begin{equation}
            \Sigma \Theta^{(i)} = \lambda^{(i)} \Theta^{(i)}
            \label{eqn:eigenvalues_vectors}
        \end{equation}
    \item The eigenvectors with the smallest eigenvalues are the best constrained components, whilst the eigenvectors with large eigenvalues are poorly constrained.
\end{enumerate}
One could therefore define an alternative to \cref{eqn:bmc} based on the number of small eigenvalues, although this itself would depend on the eigenvalue cutoff used to define ``unconstrained''.

Principle component analysis has intuitive appeal due in large part to the weight given to eigenvectors and eigenvalues early in a physicist's undergraduate mathematical education. However, in many contexts that PCA is applied, the procedure is invalid almost to the point of nonsense. 

The issue arises from the fact that the PCA procedure is not invariant under linear transformations. Typically the vectors $\theta$ have components with differing dimensionalities, in which case \eqref{eqn:eigenvalues_vectors} is dimensionally invalid.\footnote{Those that believe it is should try to answer the question: What is the dimensionality of each eigenvalue $\lambda^{(i)}$?} Equivalently, changing the units that the axes are measured in changes both the eigenvalues and eigenvectors.

For example, for \CosmoMC{} the default cosmological parameter vector is
\begin{equation}
    \theta_\mathrm{cosmo}=(\Omega_c h^2, \Omega_b h^2, 100\theta_{MC}, \tau, \log 10^{10} A_s, n_s)
    \label{eqn:theta_cosmo}
\end{equation}
the first and second components have dimensions of $10^{-4} \mathrm{km}^2\mathrm{s}^{-2}\mathrm{Mpc}^{-2}$, the third is measured in units of $10^{-2}$ radians, whilst the final three are dimensionless. 
If one were to choose a different unit/scale for any one of these (somewhat arbitrary) dimensionalities, the eigenvalues and eigenvectors would change.
To be clear, if all parameters are measured in the same units (as is the case for a traditional normal mode analysis) then PCA is a valid procedure.

Given these observations, the real question is not ``is PCA the best procedure?'', but in fact ``why does PCA usually work at all?'' The answer to this question, and an information-theoretically valid PCA will be developed in an upcoming paper.

There are two ways in which one could adjust the naive PCA procedure to be dimensionally valid. The first is simply to normalise all inputs by the prior, say by computing the prior covariance matrix
\begin{equation}
    \Sigma_0 = \left\langle (\theta-\bar\theta)(\theta-\bar\theta)^T \right\rangle_\pi, \qquad \bar{\theta} = \left\langle \theta \right\rangle_\pi
    \label{eqn:prior_cov}
\end{equation}
and then performing posterior PCA in a space normalised in some sense by this prior.

The second dimensionally valid approach would be to apply the PCA procedure to $\log \theta$. There is an implicit scale that one has to divide each component by in order to apply a logarithm, but this choice only alters the transformation by an additive constant, which PCA is in fact insensitive to. This amounts to finding components that are multiplicative combinations of parameters. A good example of such a combination is $\Omega_b h^2$, or $S_8 = \sigma_8\sqrt{\Omega_m/0.3}$, indicating that physicists are used to thinking in these terms.

\subsection{The anatomy of a Gaussian}
\begin{figure}
	\includegraphics[]{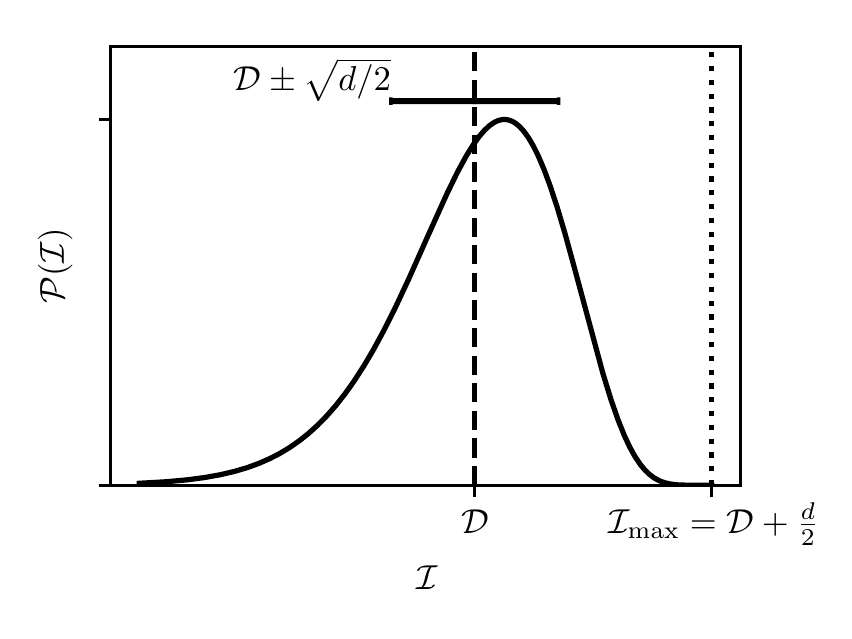}
	\caption{%
        The typical set of a $d$-dimensional Gaussian distribution can be visualised by plotting the posterior probability distribution of the Shannon information $\mathcal{I}$. The posterior has mean $\mathcal{D}$ and variance $\frac{d}{2}$. The posterior maximum occurs at $\mathcal{I}=\mathcal{D}+1$, and the domain is $(-\infty,\mathcal{I}_{\max}]$. The above plot is shown for $d=16$ in analogy with the \Planck{} likelihood from \cref{fig:shannon_examples} and \cref{tab:numerics}.\label{fig:gaussian}
	}
\end{figure}

As a concrete example of all of the above ideas, we will consider them in the context of a $d$-dimensional multivariate Gaussian. Consider a posterior $\mathcal{P}$, with parameter covariance matrix $\Sigma$ and mean $\mu$, arising from a uniform prior $\pi$ with volume $V$ which fully encompasses the posterior.  It is easy to show that the Kullback-Leibler divergence for such a distribution is
\begin{equation}
    \mathcal{D} = \log \frac{V}{\sqrt{|2\pi e \Sigma|}}
    \label{eqn:gaussian_kl}
\end{equation}
Each iso-posterior ellipsoidal contour $\mathcal{P}(\theta)=\mathcal{P}$ defines a Shannon information $\mathcal{I}=\log\mathcal{P}/\pi$. The posterior distribution $\mathcal{P}(\theta)$ induces an offset, re-scaled, $\chi^2_d$ distribution on the Shannon information
\begin{align}
    \mathcal{P}(\mathcal{I})&= \frac{1}{\Gamma(d/2)} e^{\mathcal{I}-\mathcal{I}_{\max}} {(\mathcal{I}_{\max}-\mathcal{I})}^{\frac{d}{2} - 1}, \label{eqn:chi2}\\ 
    \mathcal{I}_{\max} &= \log\frac{V}{\sqrt{|2\pi\Sigma|}} = \mathcal{D}+\frac{d}{2},\label{eqn:Imax}\\
    \mathcal{I} &\in (-\infty,\mathcal{I}_{\max}], \qquad
    \mathcal{I} \approx \mathcal{D} \pm \sqrt{d/2},
\end{align}
which may be seen graphically in \cref{fig:gaussian}.\footnote{%
    Note that in the manipulation for \cref{eqn:Imax} we have used the fact that $\log\sqrt{|2\pi e\Sigma|}=\log \sqrt{e^d|2\pi\Sigma|} = \frac{d}{2} + \log \sqrt{|2\pi\Sigma|}$.
}
This distribution has mean $\mathcal{D}$ by the definition of the Kullback-Leibler divergence, and standard deviation $\sqrt{d/2}$. The region for which the distribution $\mathcal{P}(\mathcal{I})$ is significantly non-zero defines the typical set of the posterior, indicating the Shannon information of points that would be typically drawn from the distribution $\mathcal{P}$. For this Gaussian case, the maximum posterior $\hat{\theta}_\mathrm{mp}$, likelihood $\hat{\theta}_\mathrm{ml}$ and mean $\hat{\theta}_\mathrm{m}$ parameter points coincide, and have Shannon information $I_{\max}=\mathcal{D}+\frac{d}{2}$.

\section{Bayesian model dimensionality}\label{sec:dimensionality} 

\begin{figure*}
	\includegraphics[]{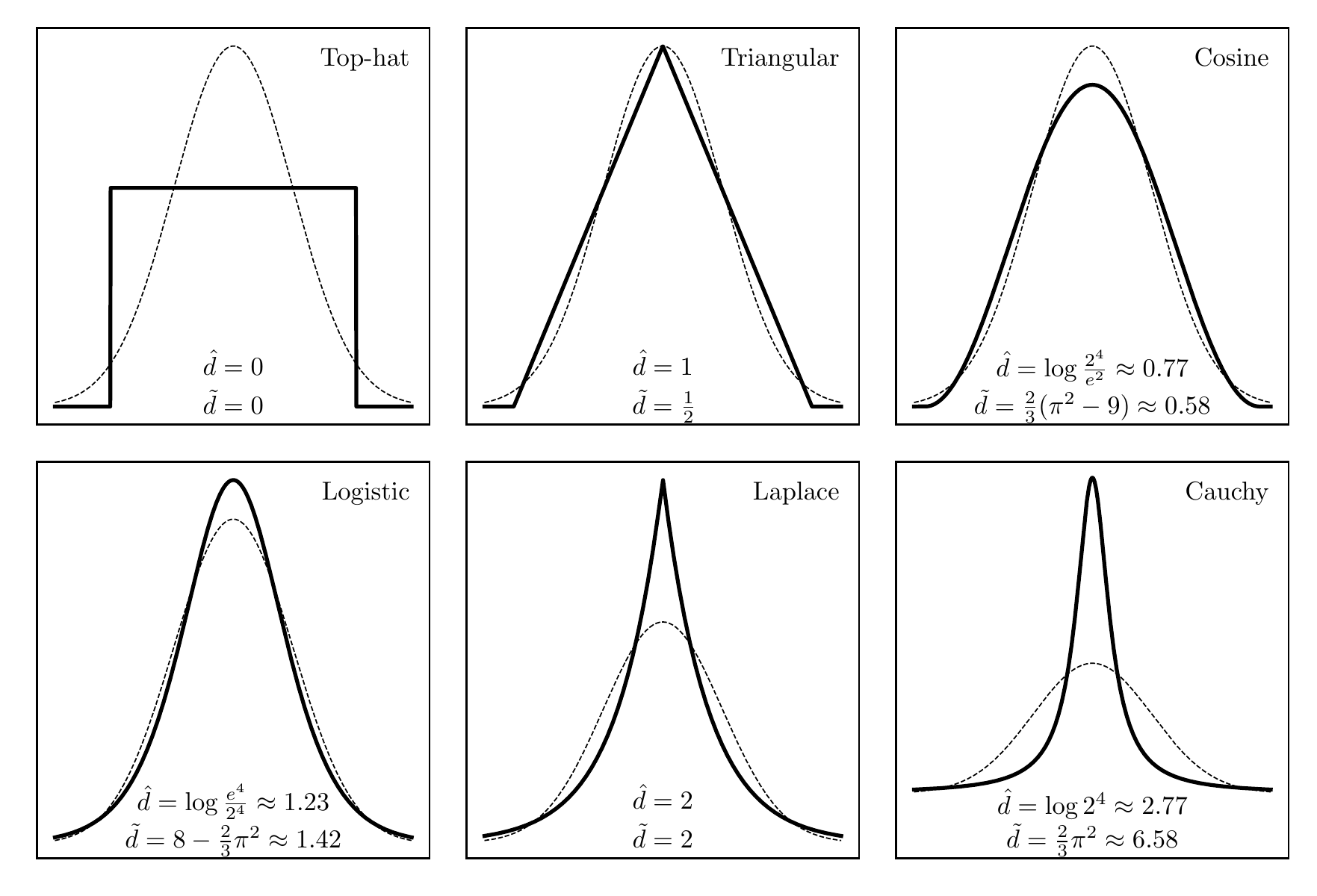}
	\caption{%
        Bayesian dimensionality for the common one-dimensional distributions in \cref{tab:analytics}. Widths are normalised so that the distributions all have the same Kullback-Leibler divergence $\mathcal{D}$. The dashed curve in all plots is a Gaussian distribution.\label{fig:examples}
	}
\end{figure*}

\subsection{The problem with Bayesian model complexity $\hat{d}$}\label{sec:bmc}

Whilst the BMC  is widely used in the statistical literature, and recovers the correct answer in the case that the posterior distribution is Gaussian, there are three key problems that should be noted.

First, it is clear that the arbitrariness regarding the choice of estimator is far from ideal, and as we shall show in \cref{sec:numerics} differing choices yield distinct and contradictory answers. A proper information theoretic quantity should be unambiguous.

Second, and most importantly in our view, estimators are not typical posterior points. In general, point estimators such as the maximum likelihood, posterior mode or mean have little statistical meaning in a Bayesian sense, since they occupy a region of vanishing posterior mass. This can be seen in \cref{fig:gaussian}, which shows that whilst an estimator may represent a point of high information, it lies in a zero posterior mass region --- If $d>2$, one can see from \cref{eqn:chi2} that $\mathcal{P}(\mathcal{I}_{\max})=0$. A physical example familiar to undergraduate quantum physicists is that of the probability distribution of an electron in a 1s orbital: The most likely location to find an electron is the origin, whilst the radial distribution function shows that the most likely region to find an electron is at the Bohr radius $a_0$.

A practical consequence of these observations is that if you choose the highest likelihood point from an MCMC chain, it will lie at a likelihood some way below the true maximum, and in general one should not expect points in the MCMC chain to lie close to the mean, mode or maximum likelihood point in likelihood space. In general, to compute these point estimators an additional calculation must be performed such as a explicit posterior and likelihood maximisation routines or a mean and likelihood computation.

Third most estimators are parameterisation dependent. Namely, if one were to transform the variables and distribution to a different coordinate system via
\begin{align}
    \theta &\to\tilde{\theta}=f(\theta),  \label{eqn:transform}\\
    \mathcal{P}(\theta)&\to \tilde{\mathcal{P}}(\tilde\theta) = \mathcal{P}(f^{-1}(\tilde\theta)) |{\partial\theta}/{\partial\tilde{\theta}}|, \label{eqn:posterior_transform}\\
    \pi(\theta)&\to \tilde{\pi}(\tilde\theta) = \pi(f^{-1}(\tilde\theta)) |{\partial\theta}/{\partial\tilde{\theta}}|,
    \label{eqn:prior_transform}
\end{align}
then neither the posterior mean from \cref{eqn:theta_mean} nor the posterior mode from \cref{eqn:theta_mp} transform under \cref{eqn:transform} if the transformation $f$ is non-linear (i.e.\ the Jacobian $|\partial\theta/\partial\tilde{\theta}|$ depends on $\tilde{\theta}$). It should be noted that this parameterisation variance is not quite as bad as it is for the PCA case, which is dependent on even linear transformations of the parameter vector.
The maximum likelihood point from \cref{eqn:theta_ml} does correctly transform, since the Jacobian terms in \cref{eqn:posterior_transform,eqn:prior_transform} cancel in the Shannon information. Parameterisation dependency is a highly undesirable ambiguity, particularly in the context of cosmology where in general the preferred choice of parameterisation varies between likelihoods and sampling codes~\cite{cosmosis,cosmomc,montepython}.

Finally, specifically to the mean estimator, for some cosmological likelihoods there may be no guarantee that the mean even lies in the posterior mass, for example in the $\sigma_8$-$\Omega_m$ banana distribution visualised by KiDS~\cite{KiDS}. In cosmology, we do not necessarily have the luxury of Gaussianity or convexity.

\subsection{The Bayesian model dimensionality $\tilde{d}$}
Considering \cref{fig:gaussian},
the fundamental concept to draw is that the BMC leverages the fact that the difference between the Shannon information $\mathcal{I}$ at the posterior peak and the mean of the posterior bulk is $d/2$ for the Gaussian case.

However, there is a second way of bringing the dimensionality out of \cref{fig:gaussian} via the variance of the posterior bulk. With this in mind, we define the {\em Bayesian model dimensionality\/} as
\begin{align}
    \frac{\tilde{d}}{2} &= \int \mathcal{P}(\theta){\left(\log\frac{\mathcal{P}(\theta)}{\pi(\theta)}-\mathcal{D}\right)}^2\d{\theta}, \\
    &= \left\langle {\mathcal{I}}^2\right\rangle_\mathcal{P} - {\left\langle \mathcal{I} \right\rangle }_\mathcal{P}^2.\label{eqn:bmd}
\end{align}
or equivalently as
\begin{equation}
    \tilde{d}/2 = \left\langle {\left(\log \mathcal{L} \right)}^2\right\rangle_\mathcal{P} - {\left\langle \log \mathcal{L} \right\rangle }^2_\mathcal{P}.
    \label{eqn:d_MCMC}
\end{equation}
We note that this form for quantifying model dimensionality is discussed in passing by \citet[p 173]{gelman} and \citet{Spiegelhalter2}, who conclude that $\tilde{d}$ is less numerically stable than $\hat{d}$. As we shall discuss in \cref{sec:numerics} we find that when applied to cosmological likelihoods the opposite is in fact true. This measure of model dimensionality is also discussed briefly in the landmark nested sampling paper by \citet{Skilling2006}, by \citet{Raveri2019}, in a cosmological context in terms of $\chi^2$ in \citet{Kunz} and \citet{information_criteria}; and  was used as part of the \Planck{} analysis~\cite{PlanckParameters2018}.

The definition of $\tilde{d}$ shares all of the desiderata that $\hat{d}$ provides, namely both $\tilde{d}$ and $\hat{d}$ are weakly prior dependent, additive for independent parameters and recover the correct answer in the Gaussian case. We believe that there are several attractive theoretical characteristics of $\tilde{d}$ that we view as advantages over $\hat{d}$

First, $\tilde{d}$ relies only on points drawn from the typical set, which is highly attractive from a Bayesian and information theoretic point of view, and more consistent when used alongside a traditional MCMC analysis of cosmological posteriors.

Second, there is a satisfying progression in the fact that whilst the mean of the Shannon information $\mathcal{D}$ gives one an overall constraint, the next order statistic (the variance) yields a measure of the dimensionality of the constraint.

Finally, in eschewing estimators this measure is completely unambiguous, as it removes all arbitrariness associated with both estimator and underlying parameterisation choice.

It should be noted that the computation of $\mathcal{D}$ requires nested sampling to provide an estimate of $\log\mathcal{Z}$. The dimensionality $\tilde{d}$ on the other hand can be computed from a more traditional MCMC chain via \cref{eqn:d_MCMC}.

\subsection{Thermodynamic interpretation}

There is a second motivation for the BMD arising from a thermodynamic viewpoint.\footnote{Historically, it was this viewpoint that drew our attention to BMD.} 
The thermodynamic generalisation of Bayes theorem is
\begin{align}
    \mathcal{L}^\beta(\theta) \times \pi(\theta) &= \mathcal{P}_\beta(\theta) \times \mathcal{Z}(\beta), 
    \label{eqn:bayes_thermo}\\
    \mathcal{Z}(\beta) &= \int \mathcal{L}^\beta(\theta) \pi(\theta) \d{\theta},
    \label{eqn:Z_thermo}
\end{align}
where on the left-hand side of \cref{eqn:bayes_thermo}, the inverse-temperature $\beta = \frac{1}{T}$ raises the likelihood $\mathcal{L}$ to the power of $\beta$ and on the right-hand side the posterior has a non-trivial dependency on temperature, denoted by a subscript $\beta$. When the evidence in \cref{eqn:bayes_thermo,eqn:Z_thermo} is a function of $\beta$ it is usually called the partition function.

The link to thermodynamics comes by considering $\theta$ to be a continuous index $i$ over microstates, the negative log-likelihood to be the energy $E$ of a microstate, and the prior to be the degeneracy of microstates $g$
\begin{gather}
    i \leftrightarrow \theta \qquad E_i \leftrightarrow - \log\mathcal{L}(\theta) \qquad g_i \leftrightarrow \pi(\theta), \nonumber\\
    g_i e^{-\beta E_i} \leftrightarrow \mathcal{L}(\theta)^\beta \pi(\theta).
    \label{eqn:thermo_correspondence}
\end{gather}
It should be noted that one of the principal advantages of nested sampling is that (other than the stopping criterion) it is blind to $\beta$, and therefore samples at all temperatures simultaneously. Nested samplers are best described as partition function calculators rather than as posterior samplers. This will be explored in further detail in an upcoming paper~\cite{aeons}.

In its thermodynamic form, the evidence becomes a generating function~\cite{generatingfunctionology}
\begin{align}
    \frac{\d{}^2}{\d{\beta}^2} \log \mathcal{Z}(\beta) &= \frac{\d{}}{\d{\beta}}\left\langle \log \mathcal{L} \right\rangle_{\mathcal{P}_\mathcal{\beta}} =  \frac{\tilde{d}}{2}, 
    \label{eqn:g2}
\end{align}
and we may identify the BMD as being related to the rate of change of average loglikelihood(energy) with inverse temperature. The BMD is therefore proportional to the Bayesian analogue of a heat capacity, $C = \frac{\d{}}{\d{T}}\langle E\rangle = \beta^2\frac{\d{}}{\d{\beta}}\langle -E\rangle$ and, like all heat capacities, is proportional to system size, or equivalently to the number of active degrees of freedom (i.e.\ dimensions).

\subsection{Analytical examples}
\begin{table}
    \begin{tabular}{|l|c|c|c|c|}
        \hline
        & $\mathcal{P}^\ast$ & $\exp(\mathcal{D}^\ast)$ & $\tilde{d}$ & $\hat{d}$\\
        \hline
        Gaussian & $e^{-x^2/2}$                  & ${1/\sqrt{2\pi e}}$ & $1$  & 1\\
        Top-hat  & $x\in[-1,1]$                    & ${1}$ & $0$  & $0$\\
        Triangle & $1-|x|$                       & ${1/\sqrt{e}}$ & $1/2$  & $1$\\
        Cosine   & $\cos^2x$ & ${e/2\pi}$ & $\frac{2(\pi^2-9)}{3} \approx 0.58$  & $\log\frac{2^4}{e^2}\approx 0.77$\\
        Logistic & $\frac{e^{-x}}{{(1+e^{-x})}^2}$       & ${1/e^2}$ & $\frac{24-2\pi^2}{3}\approx 1.42$ & $\log\frac{e^4}{2^4}\approx1.33$\\
        Laplace  & $e^{-|x|}$                    & ${1/2e}$ & $2$ & $2$\\
        Cauchy   & ${(1+x^2)}^{-1}$ & ${1/4\pi}$ & $\frac{2\pi^2}{3}\approx 6.58$  &$\log 2^4 \approx 2.77$\\
        \hline
    \end{tabular}
    \caption{Dimensionalities for one dimensional analytic distributions. The first column indicates the unnormalised probability density $\mathcal{P}^\ast(x)$. An arbitrary width $\sigma$ can be added by remapping $\mathcal{P}^\ast(x) \to \frac{1}{\sigma}\mathcal{P}^\ast(x/\sigma)$. The second column indicates the unnormalised Kullback-Leibler divergence $\mathcal{D}^\ast = \mathcal{D} - \log V/\sigma$ where the implicit prior is taken to encompass the posterior mass with width $V\gg \sigma$. The final two columns show the BMDs and BMCs respectively, which are independent of both $V$ and $\sigma$. As expected, the Gaussian has dimensionality $\tilde{d}=\hat{d}=1$, shorter and fatter distributions have lower dimensionalities, whilst narrower and taller dimensionalities have dimensions greater than one. This effect can be seen graphically in \cref{fig:examples}.\label{tab:analytics}}
\end{table}

We apply the BMD from \cref{eqn:bmd} and the BMC from \cref{eqn:bmc} to six additional univariate analytical examples: Top-hat, Triangular, Cosine, Logistic, Laplace and Cauchy. The analytical forms for the probability distribution, Kullback-Leibler divergence, BMD and BMC are listed in \cref{tab:analytics}, and plotted in \cref{fig:examples}. In all cases, we assume a uniform prior of volume $V$ which fully encompasses the posterior.

We find that whilst the Gaussian distribution gives $\tilde{d}=1$, distributions that are shorter and fatter give $\tilde{d}<1$, whilst distributions that are narrower and taller give $\tilde{d}>1$. Both measures of $\tilde{d}$ and $\hat{d}$ are in broad agreement. The Top-Hat (dimensionality $0$) and Cauchy distributions (dimensionality $\gg1$) represent pathological cases at either end of the scale, while the remainder all give dimensionalities of order $1$. In general, $\hat{d}$ is closer to unity than $\tilde{d}$, on account of the ``numerical stability'' quoted by~\citet{gelman}. However, accurate computation of $\hat{d}$ is predicated on an exact computation of the maximum, which (as shown in \cref{sec:numerics}) becomes increasingly unstable in higher dimensions and in cosmological applications. 

\begin{figure}
    \includegraphics{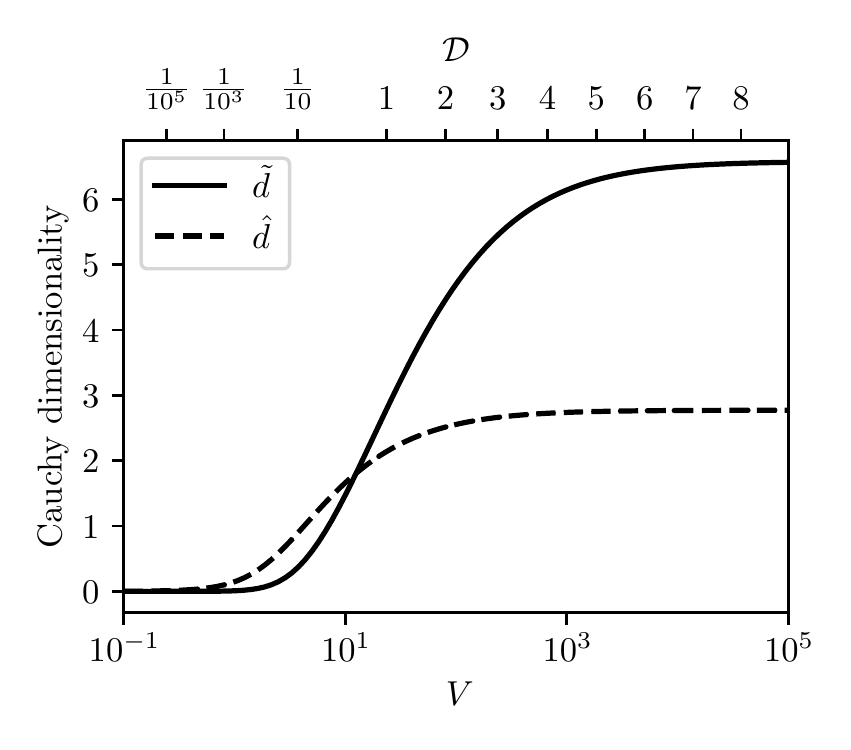}
    \caption{%
    Dependency of dimensionality and Kullback-Leibler divergence on prior volume for a Cauchy distribution $\mathcal{P}(x) \propto (1+x^2)^{-1}$. Whilst the BMD and BMC are pathologically large $(\gg 1)$ if the full domain of the Cauchy distribution is included, truncating the range to a lower prior volume $x\in [-V/2,V/2]$ reduces the dimensionality to more sensible values. \label{fig:cauchy}}
\end{figure}

It should also be noted that whilst the Cauchy distribution gives a very high dimensionality when integrated over its full infinite domain, if the domain is restricted by the prior then the dimensionality drops to more sensible values (\cref{fig:cauchy}).

\subsection{Applications}
BMDs can be used in a variety of statistical analyses. In this subsection we review a few of the possibilities.

\subsubsection{The number of constrained cosmological parameters}
As detailed in \cref{tab:numparams}, cosmological likelihoods typically introduce a large number of nuisance parameters in addition to cosmological ones, and they typically constrain a non-trivial combination of parameters. If one has datasets $A$ and $B$, one can compute the individual model dimensionalities $\tilde{d}_A$ and $\tilde{d}_B$,  as well as the model dimensionality of using both datasets together $\tilde{d}_{AB}$. Computing the crossover of these dimensionalities for any choice of $d$, $\tilde{d}$ or $\hat{d}$
\begin{equation}
    \tilde{d}_{A\cap B} = \tilde{d}_A + \tilde{d}_B - \tilde{d}_{AB}
    \label{eqn:shared_dimensionalities}
\end{equation}
will give the (effective) number of constrained cosmological parameters shared between the datasets, since any parameters constrained by just one of the datasets subtract out of the above expression. This quantity forms a cornerstone of part of the tension analysis in \citet{Tension}, and cosmological examples can be seen in the lower section of \cref{tab:numerics}.

\subsubsection{Penalising the number of model parameters}

Bayesian evidences are traditionally used in model comparison via Bayes theorem for models
\begin{equation}
    P(\mathcal{M}_i) = \frac{P(D|\mathcal{M}_i)P(\mathcal{M}_i)}{\sum_j P(D|\mathcal{M}_i)P(\mathcal{M}_j)} = \frac{\mathcal{Z}_i\Pi_i}{\sum_j\mathcal{Z}_j \Pi_j},
    \label{eqn:model_comparison}
\end{equation}
where $\Pi_i = P(\mathcal{M}_i)$ are the model priors, which are typically taken to be uniform. Often the data may not be discriminative enough to pick an unambiguously best model via the model posteriors. The correct Bayesian approach in this case is to perform model marginalisation over any future predictions \cite{model_marginalisation}. However, in other works \cite{reheating_kl,core_inflation} the Kullback-Leibler divergence has been used to split this degeneracy. The strong prior dependency of the KL divergence can make this a somewhat unfair choice for splitting this degeneracy, and users may find that the model dimensionality is a fairer choice

One implementation of this approach would be to apply a {\it post-hoc\/} model prior of
\begin{equation}
    \Pi_i(\lambda) = \lambda e^{-\lambda \tilde{d}_i}
    \label{eqn:sparse_prior}
\end{equation}
using for example $\lambda=1$. This amounts to a logarithmic Bayes factor between models of
\begin{equation}
    \log \mathcal{B}_j^i = (\log \mathcal{Z}_i - \lambda \tilde{d}_i) - (\log \mathcal{Z}_j - \lambda \tilde{d}_j)
    \label{eqn:bayes_factor}
\end{equation}
This approach is not strictly Bayesian, since $\tilde{d}_i$ is computed from the data and $\Pi_i(\lambda)$ is therefore not a true prior. However readers familiar with the concepts of sparse reconstructions~\cite{brute_force} will recognise the parallels between sparsity and this approach, as one is effectively imposing a penalty factor that promotes models that use as few parameters as necessary to constrain the data.

\subsubsection{Information criteria}
Whilst the authors' preferred method of model comparison is via the Bayesian evidence, other criteria have been used in the context of cosmology~\cite{information_criteria_0,information_criteria}: The Akaike information criterion (AIC)~\cite{AIC} and Bayesian information criterion (BIC)~\cite{BIC} are defined respectively via
\begin{align}
    \mathrm{AIC}&=-2\log\mathcal{L}_{\max} + 2k,
    \label{eqn:AIC}\\
    \mathrm{BIC}&=-2\log\mathcal{L}_{\max} + k\ln N,
    \label{eqn:BIC}
\end{align}
where $k$ is the number of parameters in the model and $N$ is the number of datapoints used in the fit. These criteria could be modified in a Bayesian sense by replacing $k$ with the BMD $\tilde{d}$. A similar modification has been discussed in the context of the deviance information criterion (DIC)~\cite{information_criteria,Kunz}.

\section{Numerical examples}\label{sec:numerics}

{%
\setlength{\tabcolsep}{0.6em}
\begin{table*}
\begin{tabular}{|c|r|r|r|r|r|r|}
\hline
Dataset                  & $\hfill\mathcal{D}_{\phantom{m}}\hfill$           & $\hfill \tilde{d}_{\phantom{m}} \hfill$                     & $\hfill \hat{d}_\mathrm{m} \hfill$          & $\hfill \hat{d}_\mathrm{mp} \hfill$         & $\hfill \hat{d}_\mathrm{ml} \hfill$ & $d$\\
\hline
\SHOES{}                 & $    2.52 \pm     0.03$ & $    0.93 \pm     0.03$ & $  -40.12 \pm     0.02$ & $    0.96 \pm     0.02$ & $    0.96 \pm     0.02$ & $6 $ \\ 
\SDSS{}                  & $    5.06 \pm     0.05$ & $    2.95 \pm     0.07$ & $   -9.55 \pm     0.05$ & $    2.93 \pm     0.05$ & $    2.93 \pm     0.05$ & $6 $ \\ 
\DES{}                   & $   22.82 \pm     0.15$ & $   14.03 \pm     0.30$ & $   10.79 \pm     0.14$ & $   14.45 \pm     0.14$ & $   17.85 \pm     0.14$ & $26$ \\ 
\Planck{}                & $   44.48 \pm     0.20$ & $   15.84 \pm     0.38$ & $   14.91 \pm     0.16$ & $   15.68 \pm     0.16$ & $   18.91 \pm     0.16$ & $21$ \\ 
\SHOES{}+\Planck{}       & $   45.02 \pm     0.20$ & $   15.97 \pm     0.36$ & $   14.64 \pm     0.15$ & $   15.39 \pm     0.15$ & $   18.40 \pm     0.15$ & $21$ \\ 
\SDSS{}+\Planck{}        & $   43.36 \pm     0.20$ & $   15.89 \pm     0.38$ & $   15.11 \pm     0.17$ & $   15.57 \pm     0.17$ & $   18.89 \pm     0.17$ & $21$ \\ 
\DES{}+\Planck{}         & $   61.13 \pm     0.25$ & $   25.88 \pm     0.62$ & $   20.79 \pm     0.25$ & $   23.54 \pm     0.25$ & $   29.30 \pm     0.25$ & $41$ \\ 
\hline                                                                                                                                                        
\SHOES{}$\cap$\Planck{}  & $    1.99 \pm     0.29$ & $    0.80 \pm     0.52$ & $  -39.84 \pm     0.23$ & $    1.25 \pm     0.23$ & $    1.48 \pm     0.23$ & $6$ \\ 
\SDSS{}$\cap$\Planck{}   & $    6.18 \pm     0.30$ & $    2.91 \pm     0.54$ & $   -9.75 \pm     0.23$ & $    3.04 \pm     0.23$ & $    2.96 \pm     0.23$ & $6$ \\ 
\DES{}$\cap$\Planck{}    & $    6.17 \pm     0.36$ & $    3.98 \pm     0.77$ & $    4.91 \pm     0.32$ & $    6.59 \pm     0.32$ & $    7.46 \pm     0.32$ & $6$ \\ 
\hline
\end{tabular}
\caption{Bayesian model dimensionalities for cosmological datasets. The first column indicates the Kullback-Leibler divergence $\mathcal{D}$ from \cref{eqn:kullback}, and the second column shows the Bayesian model dimensionality $\tilde{d}$ from \cref{eqn:bmd}. The remaining three columns show the Bayesian model complexity $\hat{d}$ from \cref{eqn:bmc} with the estimator chosen as the posterior mean, posterior mode and maximum likelihood point respectively. The final three rows show the intersection statistics, computed using the equivalents of \cref{eqn:shared_dimensionalities}.\label{tab:numerics}}
\end{table*}
}
\begin{figure}
	\includegraphics{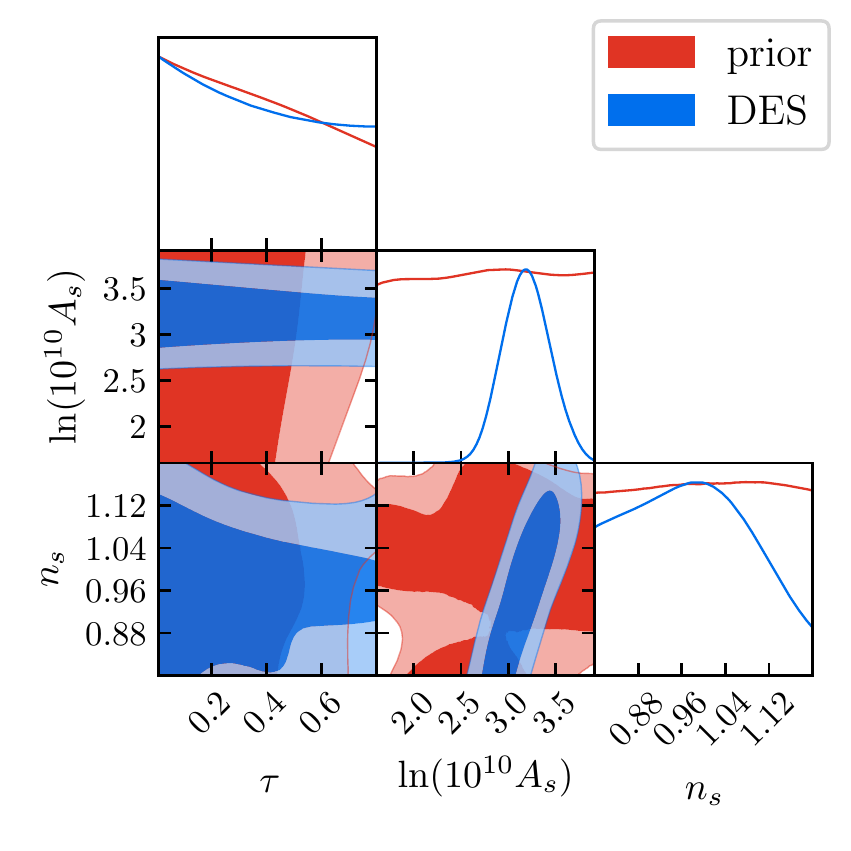}
	\caption{%
        Cosmological parameters unconstrained by \DES{}\@. Whilst \DES{} provides constraints on four of the cosmological parameters, it tells us nothing of $\tau$, and little of a correlated combination of $\ln 10^{10} A_s$ and $n_s$. This figure should be compared with \cref{fig:dimensions}. \label{fig:DES_triangle}
	}
\end{figure}

\begin{figure*}[h]
	\includegraphics[width=\textwidth]{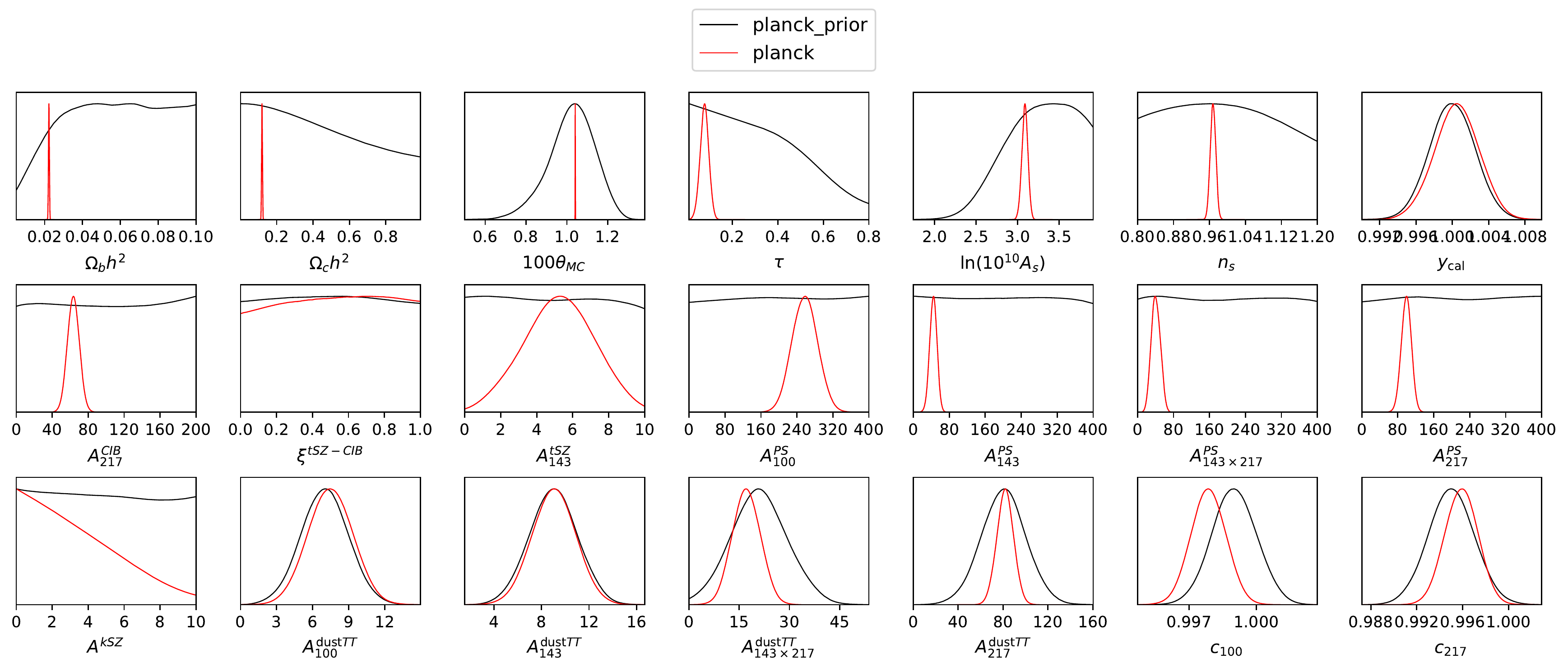}
	\caption{%
        One-dimensional marginalised default prior (black) and \Planck{} posterior (red). The Bayesian model dimensionality of $\tilde{d}_{\Planck}\approx 16$ is reflected by the fact that only a subset of the nuisance parameters are constrained by the data.\label{fig:planck_1d}
	}
\end{figure*}
\begin{figure*}
	\includegraphics[width=\textwidth]{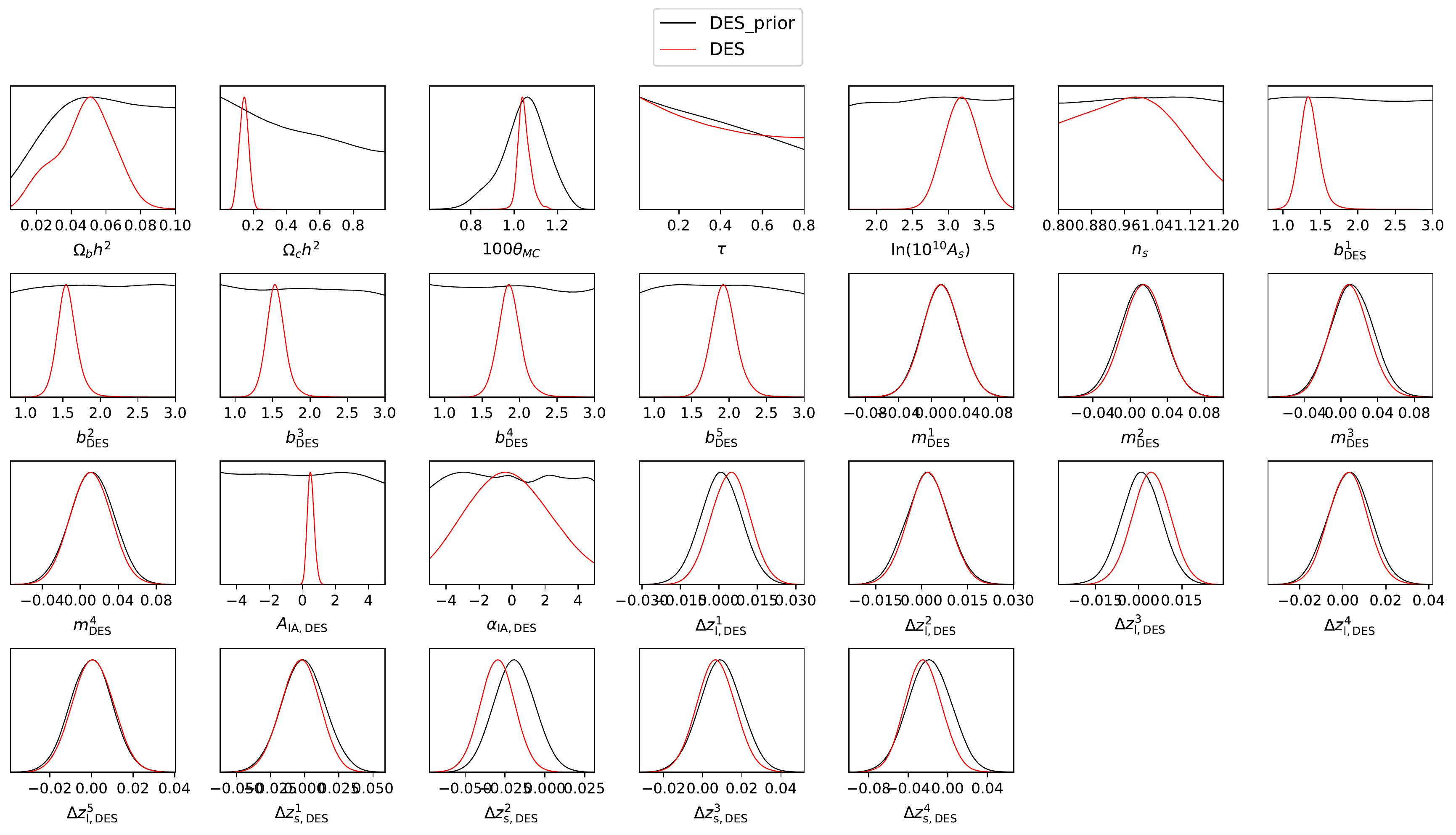}
	\caption{%
        One-dimensional marginalised default prior (black) and \DES{} Y1 posterior (red). The Bayesian model dimensionality of $\tilde{d}_\mathrm{\DES}\approx 14$ is reflected by the fact that only a combination of the cosmological parameters and a subset of the nuisance parameters are constrained by the data.\label{fig:DES_1d}
	}
\end{figure*}

\begin{figure*}
	\includegraphics[]{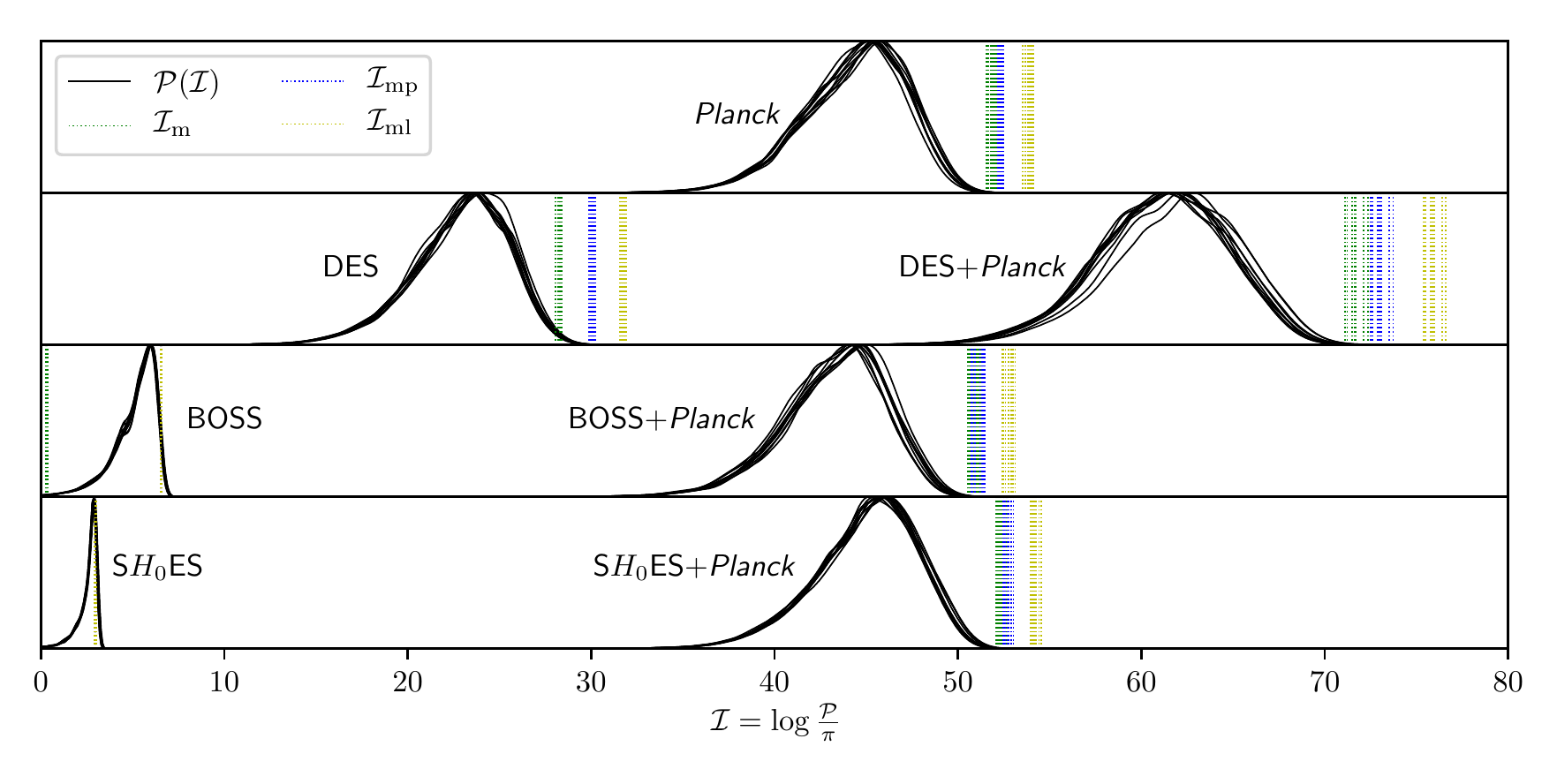}
	\caption{%
        Shannon information for the numerical examples considered in this paper. These plots are laid out in the same manner as \cref{fig:gaussian}, with the mean of each distribution representing the Kullback-Leibler divergence, and the variance the Bayesian model dimensionality. The main difference between these plots and \cref{fig:gaussian} is that the posterior mean $\mathcal{I}_\mathrm{m}$, mode $\mathcal{I}_\mathrm{mp}$ and maximum likelihood $\mathcal{I}_\mathrm{ml}$ points no longer coincide on account of the non-uniform priors and non-trivial parameterisation involved in cosmological modelling. The multiple curves for $\mathcal{P}(\mathcal{I})$ represent independent samples from the distribution of nested sampling prior volumes used to compute the Shannon information, and the spread in these curves accounts for the errors in estimating the quantities detailed in \cref{tab:numerics}.\label{fig:shannon_examples}
	}
\end{figure*}

\begin{figure*}
    \includegraphics[width=\textwidth]{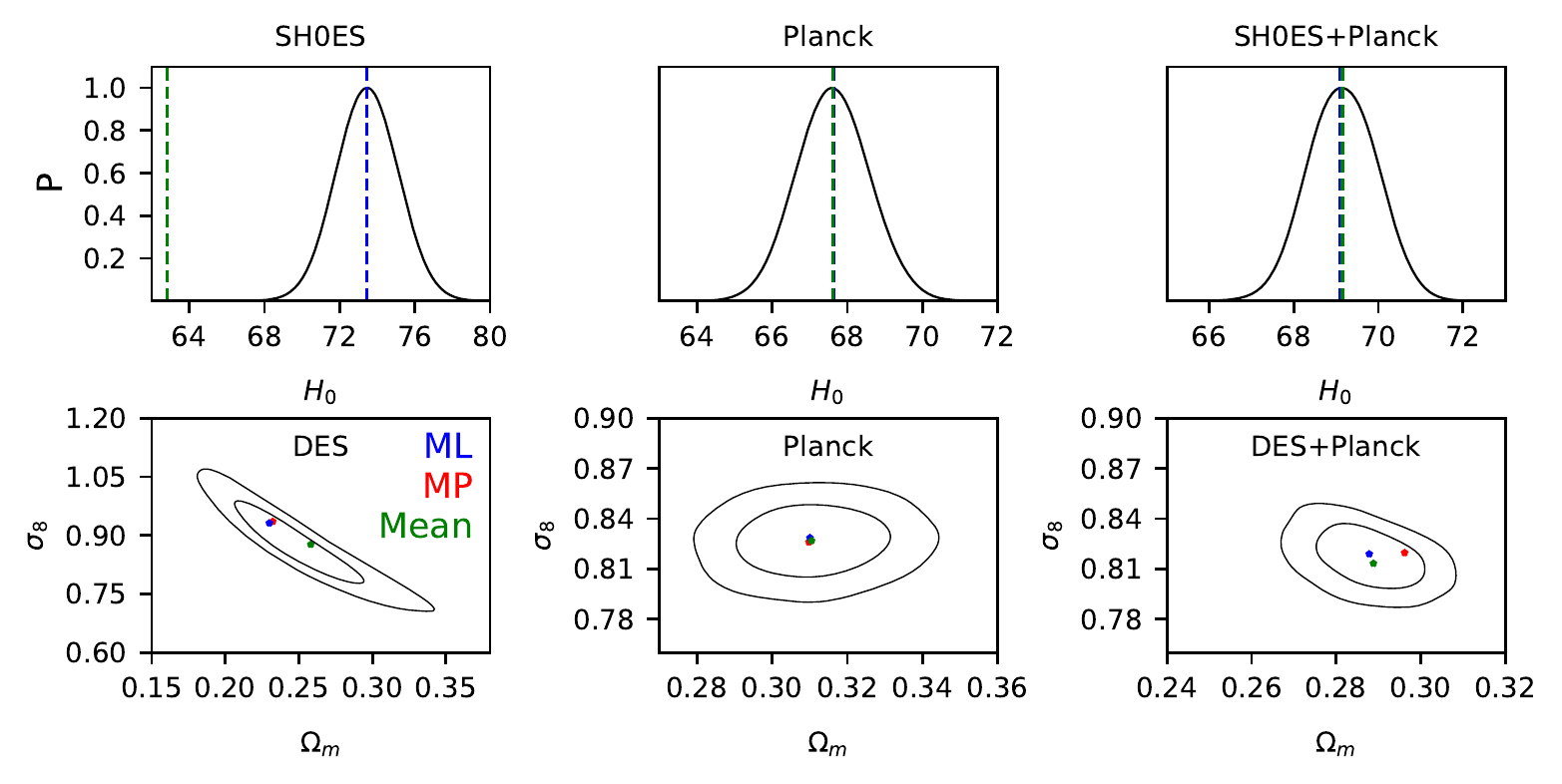}
    \caption{%
        Marginalised posterior likelihoods (in black), maximum likelihood points (ML, in blue), maximum posterior points (MP, in red) and means (in green), for some of the numerical examples used in this paper. The top plots detail the one-dimensional marginalised posterior on the Hubble parameter, whilst the lower plots show the two-dimensional marginalised posterior on $\sigma_8$ and $\Omega_m$. Top left shows the \SHOES{} likelihood, top center \Planck{}, and top right the combination of both. Bottom left shows the \DES{} posterior, bottom center \Planck{}, and bottom right their combination.\label{fig:results}
    }
\end{figure*}

\subsection{Cosmological likelihoods}\label{sec:likelihoods}

We test our method on real data by quantifying the effective number of constrained parameters in four publicly available cosmological datasets, assuming a six-parameter \LCDM cosmological model. We use the following six sampling parameters to describe this model: The density of baryonic matter $\Omega_b h^2$,  the density of cold dark matter $\Omega_c h^2$, $\theta_{MC}$ an approximation of the ratio of the sound horizon to the angular diameter distance at recombination,  the optical depth to reionisation $\tau$ and the amplitude and tilt of the primordial power spectrum $A_s$ and $n_s$. This is the default parameterisation for \CosmoMC{}~\cite{cosmomc}, and is chosen to maximize the efficiency of Metropolis-Hastings sampling codes for CMB data. The possible effects of this parameterisation choice in non-CMB constraints will be explored in future work. 

We use four key datasets in our analysis.
First, we use measurements of temperature and polarization anisotropies in the CMB measured by \Planck{} in the form of the publicly available \Planck{} 2015 data\footnote{\url{http://www.cosmos.esa.int/web/planck/pla}.}~\cite{PlanckLikelihoods2015}. 
Second, we use local cosmic distance ladder measurements of the expansion rate, using type Ia SNe calibrated by variable Cepheid stars, and implemented as a gaussian likelihood with mean and standard deviation given by the latest results obtained by the \SHOES{}\footnote{Supernovae and $H_0$ for the Equation of State.} collaboration~\cite{Riess2018}.
Third, we use the Dark Energy Survey (\DES{}) Year 1 combined analysis of cosmic shear, galaxy clustering and galaxy-galaxy lensing (a combination commonly referred to as `3x2')~\cite{DESParameters2017}.  
Finally, we use Baryon Acoustic Oscillation (BAO) measurements from the Baryon Oscillation Spectroscopic Survey (BOSS)\footnote{\url{http:// www.sdss3.org/science/BOSS_publications.php.}} DR12~\cite{SDSS,SDSS2,SDSS3}. The number of parameters that we sample over for each likelihood is described in \cref{tab:numparams}

\subsection{Nested sampling}
To compute the log-evidence $\log \mathcal{Z}$, Kullback-Leibler divergence $\mathcal{D}$ and Bayesian model dimensionality $\tilde{d}$, we use the outputs of a nested sampling run produced by \CosmoChord{}~\cite{CosmoChord,cosmomc,PolyChord0,PolyChord1,cosmomc_fs} and the equations
\begin{align}
    \mathcal{Z} \approx& \sum_{i=1}^{N}\mathcal{L}_i \times\frac{1}{2}(X_{i-1}-X_{i+1}),
    \nonumber\\
    \mathcal{D} \approx& \sum_{i=1}^{N}\frac{\mathcal{L}_i}{\mathcal{Z}}\log\frac{\mathcal{L}_i}{\mathcal{Z}} \times \frac{1}{2} (X_{i-1}-X_{i+1}),\nonumber\\
    \frac{\tilde{d}}{2} \approx& \sum_{i=1}^{N}{\frac{\mathcal{L}_i}{\mathcal{Z}}\left(\log\frac{\mathcal{L}_i}{\mathcal{Z}}-\mathcal{D}\right)}^2 \times \frac{1}{2} (X_{i-1}-X_{i+1}),\nonumber\\
    X_{i} =& t_i X_{i-1}, \qquad X_0 = 1, \qquad X_{N+1}=0,
    \nonumber\\
    P(t_i) &= n_{i} t_i^{n_i-1}\:[0<t_i<1]
    \label{eqn:simulate_t}
\end{align}
where $\mathcal{L}_i$ are the $N$ likelihood contours of the discarded points, $X_i$ are the prior volumes, $n_i$ are the number of live points and $t_i$ are real random variables. We compute $1000$ batches of the samples $\{t_i:i=1\ldots N\}$. Code for performing the above calculation is provided by the Python package \texttt{anesthetic}~\cite{anesthetic}. For our final runs, we used the \CosmoChord{} settings $n_\mathrm{live}=1000$, $n_\mathrm{prior}=10000$, with all other settings left at their defaults for \CosmoChord{} version 1.15. For more detail, see \citet{Skilling2006} or \citet{Tension}. 

In order to compute the maximum likelihood and posterior points, we found that the most reliable procedure was to use a Nelder-Mead simplex method~\cite{nelder_mead} with the initial simplex defined by the highest likelihood live points found before termination.

\subsection{Results}
Our main results are detailed in \cref{tab:numerics}, where we report the Bayesian model dimensionality $\tilde{d}$ obtained from \cref{eqn:bmd}, compared with the values obtained for the Bayesian model complexity using \cref{eqn:bmc} using three different estimators from \cref{eqn:theta_mean,eqn:theta_mp,eqn:theta_ml}: the posterior mean, posterior mode and maximum likelihood. We use the four individual datasets described in \cref{sec:likelihoods}, as well as in combination with \Planck{}. We also report the shared dimensionalities from \cref{eqn:shared_dimensionalities} using \Planck{} as the common baseline in the bottom three rows of the table.

The BMDs produce sensible values in all cases. It should be noted that in general the BMDs are lower than the number of dimensions that are sampled over (\cref{tab:numerics}): \SHOES{} constrains only one parameter ($H_0$), \SDSS{} constrains three ($\Omega_b h^2$,$\Omega_c h^2$ and a degenerate $H_0$-$A_s$ constraint), and \DES{} and \Planck{} constrain only some combinations of cosmological and nuisance parameters as shown by \cref{fig:DES_triangle,fig:planck_1d,fig:DES_1d}.

The shared dimensionalities also match cosmological intuition. For example, $\tilde{d}_\mathrm{\DES{}\cap\Planck{}}$ shows that \DES{} only constrains four cosmological parameters, as it provides no constraint on $\tau$, and only constrains a combination of $n_s$ and $\log 10^{10} A_s$. This is shown graphically in \cref{fig:DES_triangle}, which should be compared with \cref{fig:dimensions}.

All error bars on the dimensionalities arise from nested sampling's imperfect knowledge of prior volumes used to compute the posterior weights. It is likely that the error could be lowered by using a more traditional MCMC run~\cite{cosmosis,cosmomc,montepython}, although care must be taken with the MCMC error estimation since marginalising over the likelihood is numerically more unstable than that of traditional expectation values. The process of computing the Bayesian model dimensionalities and their errors is visualised in \cref{fig:shannon_examples}, which should be compared with \cref{fig:gaussian}.

The results for Bayesian model complexities on the other hand are nowhere near as satisfactory.  The arbitrariness in the choice of estimator can be seen clearly, and in general we find $\hat{d}_\mathrm{m}<\hat{d}_\mathrm{mp}<\hat{d}_\mathrm{ml}$. This variation in the choice of estimator is demonstrated graphically in \cref{fig:results}. The two maximisation estimators come out a little high, with the most extreme example being that the maximum likelihood estimator claims that there are $7.5$ shared dimensions between \DES{} and \Planck{}, which is concerning given that there are only six parameters that are shared between them. The fact that the maximum likelihood estimator consistently produces dimensionalities that are too large is unfortunate, given that it is the best motivated of all three estimators.

The mean estimator on the other hand is generally a little lower than expected and produces nonsensical results for \SHOES{} and \SDSS{} alone, where as mentioned in \cref{sec:bmc} the parameterisation variance of the estimator makes the mean extremely unreliable. In the case of \SHOES{}, we are sampling over six cosmological parameters, but only constrain $H_0$, which is only one combination of those six. As a consequence, the value of $H_0$ derived from the means of the mostly unconstrained cosmological parameters is completely prior dominated.   

The fact that \cref{fig:results} shows that the estimators are most consistent for \Planck{} data is also very telling. This is not caused by any properties of the \Planck{} data, instead it is a consequence of the parameterisation choice: All of these posteriors are obtained using the \CosmoMC{} parameterisation, which is chosen to be optimal for CMB analyses. The parameters that other surveys like \DES{} and \SHOES{} constrain are obtained as derived parameters, which changes both the mean and the maximum posterior.

\section{Conclusion}\label{sec:conclusion}

In this paper we interpret the variance in the Shannon information as a measure of Bayesian model dimensionality and present it as an alternative to Bayesian model complexity currently used in the literature. We compared these two measures of dimensionality theoretically and in the context of cosmological parameter estimation, and found that the Bayesian model dimensionality proves more accurate in reproducing results consistent with intuition.

Whilst the Bayesian model dimensionality has been acknowledged in the literature in different forms, it has yet not been widely used in cosmology. Given the ease with which the Bayesian model dimensionality can be computed from MCMC chains, we hope that this work persuades cosmologists to use this crucial statistic as a part of their analyses. For those using Nested Sampling, we hope that in future the reporting of the triple of Evidence, Kullback-Leibler divergence and Bayesian model dimensionality $(\mathcal{Z},\mathcal{D},\tilde{d})$ becomes a scientific standard.

\begin{acknowledgements}
This work was performed using resources provided by the \href{http://www.csd3.cam.ac.uk/}{Cambridge Service for Data Driven Discovery (CSD3)} operated by the University of Cambridge Research Computing Service, provided by Dell EMC and Intel using Tier-2 funding from the Engineering and Physical Sciences Research Council (capital grant EP/P020259/1), and \href{www.dirac.ac.uk}{DiRAC funding from the Science and Technology Facilities Council}.

We thank Dan Mortlock for the discussion of model dimensionalities via email correspondence following the release of \citet{Tension} which directly led to the key idea behind this paper. WH thanks Rafi Blumenfeld for lunchtime conversations on alternative PCA approaches.

WH thanks Gonville \& Caius College for their continuing support via a Research Fellowship. PL thanks STFC \& UCL for their support via a STFC Consolidated Grant.

\end{acknowledgements}

\bibliographystyle{unsrtnat}
\bibliography{D}

\begin{thebibliography}{61}
\providecommand{\natexlab}[1]{#1}
\providecommand{\url}[1]{\texttt{#1}}
\expandafter\ifx\csname urlstyle\endcsname\relax
  \providecommand{\doi}[1]{doi: #1}\else
  \providecommand{\doi}{doi: \begingroup \urlstyle{rm}\Url}\fi

\bibitem[{Scott}(2018)]{lcdm}
Douglas {Scott}.
\newblock {The Standard Model of Cosmology: A Skeptic's Guide}.
\newblock \emph{arXiv e-prints}, art. arXiv:1804.01318, Apr 2018.

\bibitem[{DES Collaboration}(2017)]{DESParameters2017}
{DES Collaboration}.
\newblock {Dark Energy Survey Year 1 Results: Cosmological Constraints from
  Galaxy Clustering and Weak Lensing}.
\newblock \emph{ArXiv e-prints}, August 2017.

\bibitem[{Planck Collaboration}(2018)]{PlanckParameters2018}
{Planck Collaboration}.
\newblock {Planck 2018 results. VI. Cosmological parameters}.
\newblock \emph{arXiv e-prints}, art. arXiv:1807.06209, July 2018.

\bibitem[{Troxel} and {Ishak}(2015)]{Troxel2015}
M.~A. {Troxel} and Mustapha {Ishak}.
\newblock {The intrinsic alignment of galaxies and its impact on weak
  gravitational lensing in an era of precision cosmology}.
\newblock \emph{\physrep}, 558:\penalty0 1--59, Feb 2015.
\newblock \doi{10.1016/j.physrep.2014.11.001}.

\bibitem[{Joachimi} et~al.(2015){Joachimi}, {Cacciato}, {Kitching}, {Leonard},
  {Mandelbaum}, {Sch{\"a}fer}, {Sif{\'o}n}, {Hoekstra}, {Kiessling}, {Kirk},
  and {Rassat}]{Joachimi2015}
Benjamin {Joachimi}, Marcello {Cacciato}, Thomas~D. {Kitching}, Adrienne
  {Leonard}, Rachel {Mandelbaum}, Bj{\"o}rn~Malte {Sch{\"a}fer}, Crist{\'o}bal
  {Sif{\'o}n}, Henk {Hoekstra}, Alina {Kiessling}, Donnacha {Kirk}, and Anais
  {Rassat}.
\newblock {Galaxy Alignments: An Overview}.
\newblock \emph{\ssr}, 193:\penalty0 1--65, Nov 2015.
\newblock \doi{10.1007/s11214-015-0177-4}.

\bibitem[{Efstathiou} and {Lemos}(2018)]{Efstathiou2018}
George {Efstathiou} and Pablo {Lemos}.
\newblock {Statistical inconsistencies in the KiDS-450 data set}.
\newblock \emph{\mnras}, 476:\penalty0 151--157, May 2018.
\newblock \doi{10.1093/mnras/sty099}.

\bibitem[{Handley} and {Lemos}(2019)]{Tension}
Will {Handley} and Pablo {Lemos}.
\newblock {Quantifying tension: interpreting the DES evidence ratio}.
\newblock \emph{arXiv e-prints}, art. arXiv:1902.04029, Feb 2019.

\bibitem[Spiegelhalter et~al.(2002)Spiegelhalter, Best, Carlin, and Van
  Der~Linde]{Spiegelhalter}
David~J. Spiegelhalter, Nicola~G. Best, Bradley~P. Carlin, and Angelika Van
  Der~Linde.
\newblock Bayesian measures of model complexity and fit.
\newblock \emph{Journal of the Royal Statistical Society Series B}, 64\penalty0
  (4):\penalty0 583--639, 2002.
\newblock URL
  \url{https://EconPapers.repec.org/RePEc:bla:jorssb:v:64:y:2002:i:4:p:583-639}.

\bibitem[Gelman et~al.(2004)Gelman, Carlin, Stern, and Rubin]{gelman}
Andrew Gelman, John~B. Carlin, Hal~S. Stern, and Donald~B. Rubin.
\newblock \emph{Bayesian Data Analysis}.
\newblock Chapman and Hall/CRC, 2nd ed. edition, 2004.

\bibitem[Spiegelhalter et~al.()Spiegelhalter, Best, Carlin, and van~der
  Linde]{Spiegelhalter2}
David~J. Spiegelhalter, Nicola~G. Best, Bradley~P. Carlin, and Angelika van~der
  Linde.
\newblock The deviance information criterion: 12 years on.
\newblock \emph{Journal of the Royal Statistical Society: Series B (Statistical
  Methodology)}, 76\penalty0 (3):\penalty0 485--493.
\newblock \doi{10.1111/rssb.12062}.
\newblock URL
  \url{https://rss.onlinelibrary.wiley.com/doi/abs/10.1111/rssb.12062}.

\bibitem[Skilling(2006)]{Skilling2006}
John Skilling.
\newblock Nested sampling for general bayesian computation.
\newblock \emph{Bayesian Anal.}, 1\penalty0 (4):\penalty0 833--859, 12 2006.
\newblock \doi{10.1214/06-BA127}.
\newblock URL \url{https://doi.org/10.1214/06-BA127}.

\bibitem[{Raveri}(2016)]{Raveri2016}
M.~{Raveri}.
\newblock {Are cosmological data sets consistent with each other within the
  {$\Lambda$} cold dark matter model?}
\newblock \emph{\prd}, 93\penalty0 (4):\penalty0 043522, February 2016.
\newblock \doi{10.1103/PhysRevD.93.043522}.

\bibitem[{Raveri} and {Hu}(2019)]{Raveri2019}
Marco {Raveri} and Wayne {Hu}.
\newblock {Concordance and discordance in cosmology}.
\newblock \emph{\prd}, 99\penalty0 (4):\penalty0 043506, Feb 2019.
\newblock \doi{10.1103/PhysRevD.99.043506}.

\bibitem[{Kunz} et~al.(2006){Kunz}, {Trotta}, and {Parkinson}]{Kunz}
Martin {Kunz}, Roberto {Trotta}, and David~R. {Parkinson}.
\newblock {Measuring the effective complexity of cosmological models}.
\newblock \emph{\prd}, 74:\penalty0 023503, Jul 2006.
\newblock \doi{10.1103/PhysRevD.74.023503}.

\bibitem[{Liddle}(2007)]{information_criteria}
Andrew~R. {Liddle}.
\newblock {Information criteria for astrophysical model selection}.
\newblock \emph{\mnras}, 377:\penalty0 L74--L78, May 2007.
\newblock \doi{10.1111/j.1745-3933.2007.00306.x}.

\bibitem[{Trotta}(2008)]{Trotta2008}
R.~{Trotta}.
\newblock {Bayes in the sky: Bayesian inference and model selection in
  cosmology}.
\newblock \emph{Contemporary Physics}, 49:\penalty0 71--104, March 2008.
\newblock \doi{10.1080/00107510802066753}.

\bibitem[MacKay(2002)]{MacKay2002}
David J.~C. MacKay.
\newblock \emph{Information Theory, Inference \& Learning Algorithms}.
\newblock Cambridge University Press, New York, NY, USA, 2002.
\newblock ISBN 0521642981.

\bibitem[Sivia and Skilling(2006)]{Sivia}
Deviderjit~Singh Sivia and John Skilling.
\newblock \emph{Data analysis : a Bayesian tutorial}.
\newblock Oxford science publications. Oxford University Press, Oxford, New
  York, 2006.
\newblock ISBN 0-19-856831-2.
\newblock URL \url{http://opac.inria.fr/record=b1133948}.

\bibitem[{Hannestad} and {Tram}(2017)]{Hannestad}
Steen {Hannestad} and Thomas {Tram}.
\newblock {Optimal prior for Bayesian inference in a constrained parameter
  space}.
\newblock \emph{arXiv e-prints}, art. arXiv:1710.08899, Oct 2017.

\bibitem[{Shannon} and {Weaver}(1949)]{Shannon:1949}
C.~E. {Shannon} and W.~{Weaver}.
\newblock \emph{{The mathematical theory of communication}}.
\newblock 1949.

\bibitem[bay(2009)]{bayesian_methods}
\emph{Bayesian Methods in Cosmology}.
\newblock Cambridge University Press, 2009.
\newblock \doi{10.1017/CBO9780511802461}.

\bibitem[Kullback and Leibler(1951)]{Kullback:1951}
S.~Kullback and R.~A. Leibler.
\newblock On information and sufficiency.
\newblock \emph{Ann. Math. Statist.}, 22\penalty0 (1):\penalty0 79--86, 03
  1951.
\newblock \doi{10.1214/aoms/1177729694}.
\newblock URL \url{https://doi.org/10.1214/aoms/1177729694}.

\bibitem[{Hosoya} et~al.(2004){Hosoya}, {Buchert}, and {Morita}]{Hoyosa:2004}
A.~{Hosoya}, T.~{Buchert}, and M.~{Morita}.
\newblock {Information Entropy in Cosmology}.
\newblock \emph{Physical Review Letters}, 92\penalty0 (14):\penalty0 141302,
  April 2004.
\newblock \doi{10.1103/PhysRevLett.92.141302}.

\bibitem[{Verde} et~al.(2013){Verde}, {Protopapas}, and {Jimenez}]{Verde:2013}
Licia {Verde}, Pavlos {Protopapas}, and Raul {Jimenez}.
\newblock {Planck and the local Universe: Quantifying the tension}.
\newblock \emph{Physics of the Dark Universe}, 2:\penalty0 166--175, September
  2013.
\newblock \doi{10.1016/j.dark.2013.09.002}.

\bibitem[{Seehars} et~al.(2014){Seehars}, {Amara}, {Refregier}, {Paranjape},
  and {Akeret}]{Seehars2014}
Sebastian {Seehars}, Adam {Amara}, Alexandre {Refregier}, Aseem {Paranjape},
  and Jo{\"e}l {Akeret}.
\newblock {Information gains from cosmic microwave background experiments}.
\newblock \emph{\prd}, 90:\penalty0 023533, July 2014.
\newblock \doi{10.1103/PhysRevD.90.023533}.

\bibitem[{Seehars} et~al.(2016){Seehars}, {Grandis}, {Amara}, and
  {Refregier}]{Seehars2016}
S.~{Seehars}, S.~{Grandis}, A.~{Amara}, and A.~{Refregier}.
\newblock {Quantifying concordance in cosmology}.
\newblock \emph{\prd}, 93\penalty0 (10):\penalty0 103507, May 2016.
\newblock \doi{10.1103/PhysRevD.93.103507}.

\bibitem[{Grandis} et~al.(2016{\natexlab{a}}){Grandis}, {Seehars}, {Refregier},
  {Amara}, and {Nicola}]{Grandis2016}
S.~{Grandis}, S.~{Seehars}, A.~{Refregier}, A.~{Amara}, and A.~{Nicola}.
\newblock {Information gains from cosmological probes}.
\newblock \emph{Journal of Cosmology and Astro-Particle Physics},
  2016:\penalty0 034, May 2016{\natexlab{a}}.
\newblock \doi{10.1088/1475-7516/2016/05/034}.

\bibitem[{Hee} et~al.(2016){Hee}, {Handley}, {Hobson}, and {Lasenby}]{HthreeL}
S.~{Hee}, W.~J. {Handley}, M.~P. {Hobson}, and A.~N. {Lasenby}.
\newblock {Bayesian model selection without evidences: application to the dark
  energy equation-of-state}.
\newblock \emph{\mnras}, 455:\penalty0 2461--2473, January 2016.
\newblock \doi{10.1093/mnras/stv2217}.

\bibitem[{Grandis} et~al.(2016{\natexlab{b}}){Grandis}, {Rapetti}, {Saro},
  {Mohr}, and {Dietrich}]{Grandis2016b}
S.~{Grandis}, D.~{Rapetti}, A.~{Saro}, J.~J. {Mohr}, and J.~P. {Dietrich}.
\newblock {Quantifying tensions between CMB and distance data sets in models
  with free curvature or lensing amplitude}.
\newblock \emph{\mnras}, 463:\penalty0 1416--1430, December 2016{\natexlab{b}}.
\newblock \doi{10.1093/mnras/stw2028}.

\bibitem[{Zhao} et~al.(2017){Zhao}, {Raveri}, {Pogosian}, {Wang}, {Crittenden},
  {Handley}, {Percival}, {Beutler}, {Brinkmann}, {Chuang}, {Cuesta},
  {Eisenstein}, {Kitaura}, {Koyama}, {L'Huillier}, {Nichol}, {Pieri},
  {Rodriguez-Torres}, {Ross}, {Rossi}, {S{\'a}nchez}, {Shafieloo}, {Tinker},
  {Tojeiro}, {Vazquez}, and {Zhang}]{Zhao2017}
Gong-Bo {Zhao}, Marco {Raveri}, Levon {Pogosian}, Yuting {Wang}, Robert~G.
  {Crittenden}, Will~J. {Handley}, Will~J. {Percival}, Florian {Beutler},
  Jonathan {Brinkmann}, Chia-Hsun {Chuang}, Antonio~J. {Cuesta}, Daniel~J.
  {Eisenstein}, Francisco-Shu {Kitaura}, Kazuya {Koyama}, Benjamin
  {L'Huillier}, Robert~C. {Nichol}, Matthew~M. {Pieri}, Sergio
  {Rodriguez-Torres}, Ashley~J. {Ross}, Graziano {Rossi}, Ariel~G.
  {S{\'a}nchez}, Arman {Shafieloo}, Jeremy~L. {Tinker}, Rita {Tojeiro}, Jose~A.
  {Vazquez}, and Hanyu {Zhang}.
\newblock {Dynamical dark energy in light of the latest observations}.
\newblock \emph{Nature Astronomy}, 1:\penalty0 627--632, August 2017.
\newblock \doi{10.1038/s41550-017-0216-z}.

\bibitem[{Nicola} et~al.(2017){Nicola}, {Amara}, and {Refregier}]{Nicola2017}
Andrina {Nicola}, Adam {Amara}, and Alexandre {Refregier}.
\newblock {Integrated cosmological probes: concordance quantified}.
\newblock \emph{Journal of Cosmology and Astro-Particle Physics},
  2017:\penalty0 045, October 2017.
\newblock \doi{10.1088/1475-7516/2017/10/045}.

\bibitem[{Nicola} et~al.(2019){Nicola}, {Amara}, and {Refregier}]{Nicola2019}
Andrina {Nicola}, Adam {Amara}, and Alexandre {Refregier}.
\newblock {Consistency tests in cosmology using relative entropy}.
\newblock \emph{Journal of Cosmology and Astro-Particle Physics},
  2019:\penalty0 011, January 2019.
\newblock \doi{10.1088/1475-7516/2019/01/011}.

\bibitem[{Lewis} and {Bridle}(2002)]{MH}
Antony {Lewis} and Sarah {Bridle}.
\newblock {Cosmological parameters from CMB and other data: A Monte Carlo
  approach}.
\newblock \emph{\prd}, 66:\penalty0 103511, Nov 2002.
\newblock \doi{10.1103/PhysRevD.66.103511}.

\bibitem[{Geman} and {Geman}(1984)]{Gibbs}
S.~{Geman} and D.~{Geman}.
\newblock Stochastic relaxation, gibbs distributions, and the bayesian
  restoration of images.
\newblock \emph{IEEE Transactions on Pattern Analysis and Machine
  Intelligence}, PAMI-6\penalty0 (6):\penalty0 721--741, Nov 1984.
\newblock ISSN 0162-8828.
\newblock \doi{10.1109/TPAMI.1984.4767596}.

\bibitem[{Betancourt}(2017)]{HMC}
Michael {Betancourt}.
\newblock {A Conceptual Introduction to Hamiltonian Monte Carlo}.
\newblock \emph{arXiv e-prints}, art. arXiv:1701.02434, Jan 2017.

\bibitem[{Planck Collaboration}(2016)]{PlanckLikelihoods2015}
{Planck Collaboration}.
\newblock {Planck 2015 results. XI. CMB power spectra, likelihoods, and
  robustness of parameters}.
\newblock \emph{\aap}, 594:\penalty0 A11, September 2016.
\newblock \doi{10.1051/0004-6361/201526926}.

\bibitem[Foreman-Mackey(2016)]{corner}
Daniel Foreman-Mackey.
\newblock corner.py: Scatterplot matrices in python.
\newblock \emph{The Journal of Open Source Software}, 24, 2016.
\newblock \doi{10.21105/joss.00024}.
\newblock URL \url{http://dx.doi.org/10.5281/zenodo.45906}.

\bibitem[{Bridges} et~al.(2009){Bridges}, {Feroz}, {Hobson}, and
  {Lasenby}]{Bridges}
M.~{Bridges}, F.~{Feroz}, M.~P. {Hobson}, and A.~N. {Lasenby}.
\newblock {Bayesian optimal reconstruction of the primordial power spectrum}.
\newblock \emph{\mnras}, 400:\penalty0 1075--1084, Dec 2009.
\newblock \doi{10.1111/j.1365-2966.2009.15525.x}.

\bibitem[{Zuntz} et~al.(2015){Zuntz}, {Paterno}, {Jennings}, {Rudd},
  {Manzotti}, {Dodelson}, {Bridle}, {Sehrish}, and {Kowalkowski}]{cosmosis}
J.~{Zuntz}, M.~{Paterno}, E.~{Jennings}, D.~{Rudd}, A.~{Manzotti},
  S.~{Dodelson}, S.~{Bridle}, S.~{Sehrish}, and J.~{Kowalkowski}.
\newblock {CosmoSIS: Modular cosmological parameter estimation}.
\newblock \emph{Astronomy and Computing}, 12:\penalty0 45--59, Sep 2015.
\newblock \doi{10.1016/j.ascom.2015.05.005}.

\bibitem[Lewis and Bridle(2002)]{cosmomc}
Antony Lewis and Sarah Bridle.
\newblock {Cosmological parameters from CMB and other data: A Monte Carlo
  approach}.
\newblock \emph{Phys. Rev.}, D66:\penalty0 103511, 2002.
\newblock \doi{10.1103/PhysRevD.66.103511}.

\bibitem[Audren et~al.(2013)Audren, Lesgourgues, Benabed, and
  Prunet]{montepython}
Benjamin Audren, Julien Lesgourgues, Karim Benabed, and Simon Prunet.
\newblock {Conservative Constraints on Early Cosmology: an illustration of the
  Monte Python cosmological parameter inference code}.
\newblock \emph{JCAP}, 1302:\penalty0 001, 2013.
\newblock \doi{10.1088/1475-7516/2013/02/001}.

\bibitem[{van Uitert} et~al.(2018){van Uitert}, {Joachimi}, {Joudaki}, {Amon},
  {Heymans}, {K{\"o}hlinger}, {Asgari}, {Blake}, {Choi}, {Erben}, {Farrow},
  {Harnois-D{\'e}raps}, {Hildebrandt}, {Hoekstra}, {Kitching}, {Klaes},
  {Kuijken}, {Merten}, {Miller}, {Nakajima}, {Schneider}, {Valentijn}, and
  {Viola}]{KiDS}
Edo {van Uitert}, Benjamin {Joachimi}, Shahab {Joudaki}, Alexandra {Amon},
  Catherine {Heymans}, Fabian {K{\"o}hlinger}, Marika {Asgari}, Chris {Blake},
  Ami {Choi}, Thomas {Erben}, Daniel~J. {Farrow}, Joachim {Harnois-D{\'e}raps},
  Hendrik {Hildebrandt}, Henk {Hoekstra}, Thomas~D. {Kitching}, Dominik
  {Klaes}, Konrad {Kuijken}, Julian {Merten}, Lance {Miller}, Reiko {Nakajima},
  Peter {Schneider}, Edwin {Valentijn}, and Massimo {Viola}.
\newblock {KiDS+GAMA: cosmology constraints from a joint analysis of cosmic
  shear, galaxy-galaxy lensing, and angular clustering}.
\newblock \emph{\mnras}, 476:\penalty0 4662--4689, Jun 2018.
\newblock \doi{10.1093/mnras/sty551}.

\bibitem[{Handley} et~al.(2019)]{aeons}
Will {Handley} et~al.
\newblock \emph{AEONS: Approximate end of nested sampling}.
\newblock 2019.

\bibitem[Wilf(2006)]{generatingfunctionology}
Herbert~S. Wilf.
\newblock \emph{Generatingfunctionology}.
\newblock A. K. Peters, Ltd., Natick, MA, USA, 2006.
\newblock ISBN 1568812795.

\bibitem[{Gariazzo} and {Mena}(2019)]{model_marginalisation}
S.~{Gariazzo} and O.~{Mena}.
\newblock {Cosmology-marginalized approaches in Bayesian model comparison: The
  neutrino mass as a case study}.
\newblock \emph{\prd}, 99:\penalty0 021301, Jan 2019.
\newblock \doi{10.1103/PhysRevD.99.021301}.

\bibitem[{Martin} et~al.(2016){Martin}, {Ringeval}, and {Vennin}]{reheating_kl}
J{\'e}r{\^o}me {Martin}, Christophe {Ringeval}, and Vincent {Vennin}.
\newblock {Information gain on reheating: The one bit milestone}.
\newblock \emph{\prd}, 93:\penalty0 103532, May 2016.
\newblock \doi{10.1103/PhysRevD.93.103532}.

\bibitem[Collaboration(2018)]{core_inflation}
CORE Collaboration.
\newblock {Exploring cosmic origins with CORE: Inflation}.
\newblock \emph{Journal of Cosmology and Astro-Particle Physics},
  2018:\penalty0 016, Apr 2018.
\newblock \doi{10.1088/1475-7516/2018/04/016}.

\bibitem[{Higson} et~al.(2019){Higson}, {Handley}, {Hobson}, and
  {Lasenby}]{brute_force}
Edward {Higson}, Will {Handley}, Michael {Hobson}, and Anthony {Lasenby}.
\newblock {Bayesian sparse reconstruction: a brute-force approach to
  astronomical imaging and machine learning}.
\newblock \emph{\mnras}, 483:\penalty0 4828--4846, Mar 2019.
\newblock \doi{10.1093/mnras/sty3307}.

\bibitem[{Liddle}(2004)]{information_criteria_0}
Andrew~R. {Liddle}.
\newblock {How many cosmological parameters?}
\newblock \emph{\mnras}, 351:\penalty0 L49--L53, Jul 2004.
\newblock \doi{10.1111/j.1365-2966.2004.08033.x}.

\bibitem[{Akaike}(1974)]{AIC}
H.~{Akaike}.
\newblock {A New Look at the Statistical Model Identification}.
\newblock \emph{IEEE Transactions on Automatic Control}, 19:\penalty0 716--723,
  Jan 1974.

\bibitem[{Schwarz}(1978)]{BIC}
Gideon {Schwarz}.
\newblock {Estimating the Dimension of a Model}.
\newblock \emph{Annals of Statistics}, 6:\penalty0 461--464, Jul 1978.

\bibitem[{Riess} et~al.(2018){Riess}, {Casertano}, {Yuan}, {Macri}, {Anderson},
  {MacKenty}, {Bowers}, {Clubb}, {Filippenko}, {Jones}, and
  {Tucker}]{Riess2018}
Adam~G. {Riess}, Stefano {Casertano}, Wenlong {Yuan}, Lucas {Macri}, Jay
  {Anderson}, John~W. {MacKenty}, J.~Bradley {Bowers}, Kelsey~I. {Clubb},
  Alexei~V. {Filippenko}, David~O. {Jones}, and Brad~E. {Tucker}.
\newblock {New Parallaxes of Galactic Cepheids from Spatially Scanning the
  Hubble Space Telescope: Implications for the Hubble Constant}.
\newblock \emph{\apj}, 855:\penalty0 136, March 2018.
\newblock \doi{10.3847/1538-4357/aaadb7}.

\bibitem[{Alam} and et~al.(2017)]{SDSS}
Shadab {Alam} and et~al.
\newblock {The clustering of galaxies in the completed SDSS-III Baryon
  Oscillation Spectroscopic Survey: cosmological analysis of the DR12 galaxy
  sample}.
\newblock \emph{\mnras}, 470:\penalty0 2617--2652, September 2017.
\newblock \doi{10.1093/mnras/stx721}.

\bibitem[{Beutler} et~al.(2011){Beutler}, {Blake}, {Colless}, {Jones},
  {Staveley-Smith}, {Campbell}, {Parker}, {Saunders}, and {Watson}]{SDSS2}
Florian {Beutler}, Chris {Blake}, Matthew {Colless}, D.~Heath {Jones}, Lister
  {Staveley-Smith}, Lachlan {Campbell}, Quentin {Parker}, Will {Saunders}, and
  Fred {Watson}.
\newblock {The 6dF Galaxy Survey: baryon acoustic oscillations and the local
  Hubble constant}.
\newblock \emph{\mnras}, 416:\penalty0 3017--3032, October 2011.
\newblock \doi{10.1111/j.1365-2966.2011.19250.x}.

\bibitem[{Ross} et~al.(2015){Ross}, {Samushia}, {Howlett}, {Percival},
  {Burden}, and {Manera}]{SDSS3}
Ashley~J. {Ross}, Lado {Samushia}, Cullan {Howlett}, Will~J. {Percival}, Angela
  {Burden}, and Marc {Manera}.
\newblock {The clustering of the SDSS DR7 main Galaxy sample - I. A 4 per cent
  distance measure at z = 0.15}.
\newblock \emph{\mnras}, 449:\penalty0 835--847, May 2015.
\newblock \doi{10.1093/mnras/stv154}.

\bibitem[{Handley}(2019)]{CosmoChord}
W.~J. {Handley}.
\newblock Cosmochord 1.15, January 2019.
\newblock URL \url{https://doi.org/10.5281/zenodo.2552056}.

\bibitem[{Handley} et~al.(2015{\natexlab{a}}){Handley}, {Hobson}, and
  {Lasenby}]{PolyChord0}
W.~J. {Handley}, M.~P. {Hobson}, and A.~N. {Lasenby}.
\newblock {POLYCHORD: nested sampling for cosmology}.
\newblock \emph{\mnras}, 450:\penalty0 L61--L65, June 2015{\natexlab{a}}.
\newblock \doi{10.1093/mnrasl/slv047}.

\bibitem[{Handley} et~al.(2015{\natexlab{b}}){Handley}, {Hobson}, and
  {Lasenby}]{PolyChord1}
W.~J. {Handley}, M.~P. {Hobson}, and A.~N. {Lasenby}.
\newblock {POLYCHORD: next-generation nested sampling}.
\newblock \emph{\mnras}, 453:\penalty0 4384--4398, November 2015{\natexlab{b}}.
\newblock \doi{10.1093/mnras/stv1911}.

\bibitem[Lewis(2013)]{cosmomc_fs}
Antony Lewis.
\newblock {Efficient sampling of fast and slow cosmological parameters}.
\newblock \emph{Phys. Rev.}, D87:\penalty0 103529, 2013.
\newblock \doi{10.1103/PhysRevD.87.103529}.

\bibitem[Handley(2019)]{anesthetic}
Will Handley.
\newblock anesthetic: nested sampling visualisation.
\newblock \emph{The Journal of Open Source Software}, 4\penalty0 (37), Jun
  2019.
\newblock \doi{10.21105/joss.01414}.
\newblock URL \url{http://dx.doi.org/10.21105/joss.01414}.

\bibitem[Nelder and Mead(1965)]{nelder_mead}
J.~A. Nelder and R.~Mead.
\newblock {A Simplex Method for Function Minimization}.
\newblock \emph{The Computer Journal}, 7\penalty0 (4):\penalty0 308--313, 01
  1965.
\newblock ISSN 0010-4620.
\newblock \doi{10.1093/comjnl/7.4.308}.
\newblock URL \url{https://dx.doi.org/10.1093/comjnl/7.4.308}.

\end{thebibliography}

\end{document}